\documentclass[letterpaper,titlepage,11pt]{article}
\usepackage{latexsym,amssymb,amstext,amsmath}
\usepackage{slashed}
\usepackage{mathrsfs}
\usepackage{color}
\usepackage[svgnames]{xcolor}
\usepackage{enumerate}
\usepackage{graphicx}
\usepackage{tikz}
\usepackage{latexsym}
\usepackage{cite}
\usepackage{caption}
\usepackage[symbol]{footmisc}
\usepackage{mathtools}
\usepackage{enumitem}

\allowdisplaybreaks

\usepackage[
colorlinks=true,
linkcolor=DarkBlue,
anchorcolor=Black,
urlcolor=DarkRed,
filecolor=green,
citecolor=DarkRed,
linktoc=page,
pdfstartview=FitV,
pdftitle={(Super)Universal Attractors and the de Sitter Vacua in String Landscape},
pdfauthor={Omer Guleryuz},
pdfsubject={(Super)Universal Attractors and the de Sitter Vacua in String Landscape},
pdfkeywords={Supergravity, Inflation, Universal Attractors, de Sitter Vacua, String Landscape},
bookmarksopen=true
]{hyperref}

	\makeatletter
	\@addtoreset{equation}{section}
	\makeatother
	
\begin{document}
		%
		\begin{titlepage}
			
			\bigskip
			
			\begin{center}
				{\LARGE\bfseries (Super)Universal Attractors and the de Sitter Vacua in String Landscape}
				\\[10mm]
				\textbf{Omer Guleryuz$^{1,2}$}\\[5mm]
				\vskip 25pt

				{\em  \hskip -.1truecm $^1$Department of Physics, Istanbul Technical University,  \\
					Maslak 34469 Istanbul, Turkey  \vskip 5pt }
					
				{\em  \hskip -.1truecm $^2$Jefferson Physical Laboratory, Harvard University,  \\
					Cambridge, MA 02138, USA  \vskip 5pt }
				
				{E-mail: {\tt \href{mailto:omerguleryuz@itu.edu.tr}{omerguleryuz@itu.edu.tr}}}

			\end{center}
			
			\vspace{3ex}

			\begin{center}
				{\bfseries Abstract}
			\end{center}
			\begin{quotation} 
			In this work, we present an effective field theory for string inflation with spontaneously broken supersymmetry without generating any supersymmetric anti-de Sitter vacua. In that regard, we analyze the nilpotent superfields that effectively capture the physics of anti-D3 branes, and obtain the underlying pattern of universal attractors with a single parameter. Accordingly, we reveal a novel uplifting method by adding the same parameter as a complex contribution parallel to the decomposition of a superfield. Following that, we obtain an almost vanishing cosmological constant in a region where the inflationary attractors unify. Finally, we show that the introduction of nilpotent superfields drastically extends the string landscape for the de Sitter (swampland) conjecture, and the (super)universal attractors are in the string landscape in that respect.
			\end{quotation}
			
			\vfill
			
            \begin{center}{\flushleft{\today}}\end{center}
		\end{titlepage}
		\setcounter{page}{1}
		\tableofcontents

\newpage
\section{Introduction}
\hspace{4mm}  In the beginning, the efforts of constructing a de Sitter (dS) vacuum in string theory, known as the famous \textit{KKLT method} \cite{Kachru:2003aw}, led to the realization that; $\overline{D3}$ branes in an $\mathcal{N}=1$, $D=4$ background can be interpreted as equivalent to a theory of nilpotent supergravity (SUGRA) with spontaneously broken supersymmetry (SUSY) \cite{Bergshoeff:2015jxa,Kallosh:2014wsa}. Initially, such a connection with constrained superfields and super-Dp-branes \cite{Cederwall:1996pv,Cederwall:1996ri,Bergshoeff:1996tu,Aganagic:1996nn,Aganagic:1996pe} was studied in the context of non-linear realization of SUSY within Volkov-Akulov goldstino theory \cite{Volkov:1972jx,Volkov:1973ix} in \cite{Rocek:1978nb,Ivanov:1978mx,PhysRevD.19.2300,Casalbuoni:1988xh,Rocek:1997hi,Komargodski:2009rz,Antoniadis:2010hs,Kuzenko:2010ef,Ferrara:2016een}. In string theory context, these constrained superfields very efficiently capture the physics of $\overline{D3}$ branes at the tip of a Calabi-Yau throat and one can provide a connection between SUGRA and string theory, in case one considers constrained superfields (i.e. $X^2=0$) as also studied in \cite{Kallosh:2015sea,Kallosh:2015tea,Aparicio:2015psl,Bandos:2016xyu,Bertolini:2015hua,Dasgupta:2016prs,Garcia-Etxebarria:2015lif,Kallosh:2015nia,Dudas:2015eha,Kallosh:2016aep,Vercnocke:2016fbt,Burgess:2022nbx,DallAgata:2022abm,Kallosh:2022fsc}. Here the constrained superfields fulfill the nilpotency condition of the nilpotent SUGRA. That constraint also gives rise to many simplifications and advantages in cosmic evolution scenarios constructed by nilpotent superfields, including the inflation theory (see e.g.  \cite{Antoniadis:2014oya,Ferrara:2014kva,Ferrara:2015tyn,Scalisi:2015qga,Carrasco:2015pla,Carrasco:2015iij,DallAgata:2014qsj,Kallosh:2014via,Kallosh:2014hxa,Linde:2016bcz,McDonough:2016der,Kallosh:2017wnt,Antoniadis:2018oeh}). One can categorize these advantages as follows;
\begin{itemize}
    \item[\textbf{i.}] It opens the possibility of uplifting SUGRA potentials in the inflation context, analogous to $\alpha'$ corrections to the Kähler potential as occurs in \textit{Kähler uplifting method} \cite{Louis:2012nb,Westphal:2006tn}.
    \item[\textbf{ii.}] It provides a better framework for controlling and overall improving stability conditions for the corresponding model since the nilpotent superfields are constrained, and the scalar part of the nilpotent superfield is built from bilinear fermions (hence has zero vacuum expectation value (vev)).
    \item[\textbf{iii.}] One can also decouple inflationary dynamics with nilpotent superfields considering the flat directions \cite{McDonough:2016der}. This procedure has been further analyzed with some extra conditions in \cite{Kallosh:2021vcf} for potentials derived from type IIB string theory and called the \textit{Sequestered Inflation} \cite{Kallosh:2021fvz}.
\end{itemize}

Following this prospects; string theory and $\mathcal{N}=1$, $D=4$ SUGRA provide a rich landscape of mechanisms for bridging high-energy (UV) and low-energy (IR) physics, SUSY breaking, dS vacuum, dark energy and inflation. Consequently, this bridge is a testing ground for the cosmological observables of the current paradigm and their high energy sensitivity towards inflation in string theory\cite{Kachru:2003sx}.

In the first part of this study, we examine classifications of general inflation potentials, spontaneous SUSY breaking, and uplifting mechanisms via their connections to the nilpotent Kähler corrections from an effective field theory (EFT) perspective. In section \ref{Sec2.1}, we employed the Sequestered inflation model as a toy model, which considers only the bilinear nilpotent couplings. After analyzing the cosmological inflation models in this context, we denoted the issues regarding the fine-tuning of the bilinear nilpotent couplings and the corresponding uplifting mechanism. Then in section \ref{Sec2.2}, we revealed the whole picture by including linear nilpotent couplings and derived the most general form of the SUGRA potential. Finally, we covered various types of linear nilpotent couplings to investigate the resulting modifications in inflationary dynamics, spontaneous SUSY breaking, and the uplifting mechanism. That analysis also presented us with some signs of an escape from the issues of the original KKLT uplifting method\footnote{In a recent study of holography and the KKLT scenario \cite{Lust:2022lfc}, it has been argued that the construction of a supersymmetric anti-dS (AdS) vacuum as in the first step of the KKLT is beyond the validity of the EFT.} via additional linear nilpotent coupling contributions in the F-terms.

For the second part of this study, we first briefly reviewed the universal attractors for inflation \cite{Kallosh:2013tua} and pointed out the underlying pattern that emerges in kinetic and potential terms in section \ref{SUAFNC}. We showed that nilpotent SUGRA is a natural candidate to capture the same pattern. Moreover, embedding universal attractors to nilpotent SUGRA allowed us to sequester the scalar potential and the kinetic term, thus obtaining the potential with nilpotent couplings and the kinetic term with the remainder of the Kähler potential separately. Consequently, it also expanded the suitable Kähler potential space for obtaining the same kinetic term as the one in the universal attractor mechanism. Following that in section \ref{Sec3.1}, we produced an uplifted potential without generating any SUSY AdS vacua and established an explicit dependence of inflation with $\overline{D3}$ brane contribution, or equivalently with the gravitino mass. In that regard, we analyzed a unified region for inflationary attractors where the dS vacua for the (super)universal attractors appear naturally for an almost vanishing cosmological constant in section \ref{Sec3.2}. In section \ref{Sec3.3}, we showed that the moduli are stable and respect the expected mass hierarchy for all values of the canonically normalized real fields during the inflation. Then, in section \ref{Sec3.4}, we embedded the nilpotent superfields to the dS Conjecture (dSC) \cite{Obied:2018sgi,Garg:2018reu,Ooguri:2018wrx}. We showed that the resulting string landscape drastically expands for the consistent EFT space with that conjecture. Finally, we denoted our results and concluding remarks in section \ref{Sec4}. 

We also discussed two key equivalences in Appendix \ref{Appendix}. We first point out a canonical inflation case with multiple fields in \ref{A.1}, where the slow-roll parameter becomes approximately equivalent to the potential gradient flow. And we show that this model class is generally excluded by dSC. Then in \ref{A.2}, we emphasize a particular transition between the generic superpotential and Kähler potential terms coupled with nilpotent superfields. We show that this transition produces an equivalence connecting two main classes of models while preserving the properties of nilpotent SUGRA.



\section{Generic Scalar Potentials and Sequestered Inflation}\label{FirstPart}
\hspace{4mm}
In this section, we will classify general inflation potentials in the context of SUGRA and string theory. It is appropriate to consider the SUGRA potentials with two main generating sectors: the Kähler potential and the superpotential. Phenomenological inflationary potentials may get a contribution from both sectors. Here, we will examine cases where the generic scalar potentials emerge only from the Kähler potential with nilpotent corrections. In that regard, we will first review the bilinear nilpotent superfields with generic coupling terms as a toy model and analyze the resulting dynamics from a cosmological perspective.

\subsection{Bilinear Nilpotent Corrections}\label{Sec2.1}
\hspace{4mm} In string theory context (as a backreaction), an $\overline{D3}$ brane on a bulk geometry would translate into nilpotent superfields coupled to Kähler moduli in $4D$ SUGRA. As firstly denoted in \cite{McDonough:2016der}, following the same steps, one can develop a class of cosmological SUGRA models where the inflaton appears only as a part of the nilpotent Kähler potential corrections. These possible corrections appear in the Kähler potential with terms as either bilinear or linear in $X$ and $\bar{X}$ as
\begin{equation}
    K (\Phi, \bar{\Phi}, X, \bar{X})= ... + f(\Phi,\Bar{\Phi}) X \Bar{X} + g(\Phi,\Bar{\Phi}) X + \bar{g}(\Phi,\Bar{\Phi}) \Bar{X}
\end{equation}
where $X$ is a nilpotent superfield, $f(\Phi,\Bar{\Phi})$ and $g(\Phi,\Bar{\Phi})$ are non-holomorphic arbitrary functions of the bulk modulus. Since these couplings are non-holomorphic, they cannot be disappeared away with a Kähler transformation. One can find further motivations of this class of couplings considering that D-brane matter fields also appear with their complex conjugate in string theory. Moving on, in order to allow for an extremum along with the flat directions $\Phi = \Bar{\Phi}$, imposing the reality condition via the symmetry $\Phi \rightarrow \bar{\Phi}$ would be sufficient. This symmetry also allows one to define a function
\begin{equation}
    f(\Phi,\Bar{\Phi}) \equiv \frac{1}{V_{\text{inf}}(\Phi)},
\end{equation}
along the same trajectory. Then for a simple example \cite{McDonough:2016der}, given the Kähler potential
\begin{equation}\label{FirstKählerPot}
    K(\Phi, \bar{\Phi}, X, \bar{X})=-\frac{1}{2}\left(\Phi-\Bar{\Phi}\right)^{2} + \frac{1}{V_{\text{inf}}(\Phi)} X \Bar{X}
\end{equation}
and the superpotential, which describes a dS phase of SUGRA \cite{Bergshoeff:2015tra,Schillo:2015ssx,Hasegawa:2015bza,Bergshoeff:2016psz},
\begin{equation}
    W (X)=W_0 + M X,
\end{equation}
the SUGRA potential takes the form of
\begin{equation}\label{FirstSUGRAPot}
    V(\Phi)=  M^2 V_{\text{inf}}(\Phi)-3 W_{0}^{2}
\end{equation}
at the inflationary trajectory $(X=\Phi - \Bar{\Phi} = 0)$. The string theory encoding is established with the $\overline{D3}$ brane embedded in the nilpotent superfield $X$ and the bulk geometry (or fluxes) embedded in $\Phi$. Therefore, $M$ parameterizes the $\overline{D3}$ brane contribution (SUSY breaking scale) and the gravitino mass $m_{3/2} = W_0$ appears as the flux-induced superpotential \cite{Giddings:2001yu}. Here the F-term SUGRA potential is calculated with
\begin{equation}
    V_{\text{SUGRA}} = \rm{e}^{\textit{K}} \left[\left(\mathcal{D}W\right)^2 -3 W \overline{W} \right],
\end{equation}
where we take $M_{pl}=1$ for simplicity and $\mathcal{D}_{\alpha} W = \partial_{\alpha} W + \left(\partial_{\alpha} K\right) W$ is the Kähler covariant derivative. If one only considers the bilinear nilpotent Kähler corrections as in this case, then the corresponding F-terms become
\begin{equation}
    \mathcal{D}_{\Phi} W =0 \;\;\; \text{and} \;\;\;  \mathcal{D}_{X} W = M,
\end{equation}
and the SUSY is spontaneously and purely broken along the $X$-direction. One should also note here that; in case the first term of the Kähler potential (\ref{FirstKählerPot}), was not shift symmetric in $\Phi$, one can apply a Kähler transformation that would be sufficient to give it a shift symmetry. Therefore the inflationary trajectory would be protected for the flat directions $\Phi = \Bar{\Phi}$ and which is also a crucial property that will be discussed in section \ref{SUAFNC}. 

The SUGRA potential (\ref{FirstSUGRAPot}) contains an arbitrary scalar potential for inflation, and one can interpret it as a residual cosmological constant in case the inflation potential reaches its minimum at late times of cosmic evolution. In general, the following two situations can be considered for this potential;
\begin{itemize}
\item[\textbf{i.}] For a spontaneous symmetry breaking potential, i.e. Higgs like potentials, it would reach its minimum at the vev $v$ as $V_{\text{inf}}(\phi = v)$.
\item[\textbf{ii.}]  In case the inflation potential is not a symmetry-breaking type, to obtain a positive cosmological constant, it should reach a minimum value away from zero.
\end{itemize}
It is clear to see for both situations, to obtain a cosmological scenario with a positive cosmological constant, the scalar part of the inflation potential should not disappear since it requires that
\begin{equation}
    \Lambda=  M^2 V_{\text{inf}}(\Phi)-3 W_{0}^{2} > 0
\end{equation}
at the minimum of the potential. Another issue is that in case one removes the $\overline{D3}$ brane contribution by taking $M\rightarrow 0$, SUSY is restored and one obtains a pure AdS state. This corresponds to the first step of the KKLT uplifting method, which is argued to be ruled out with a holography argument in a recent study \cite{Lust:2022lfc}.
\vspace{-3mm}
\paragraph{Decoupling inflationary dynamics:}
So far, considering nilpotent superfields in the context of string inflation set us a very practical cosmological background. With a similar methodology, it is possible to take string inflation one step further and use Kähler potentials directly derived from string theory without affecting the inflation potential as described in the sequestered inflation method \cite{Kallosh:2021vcf}. One can generally denote the corresponding setup with
\begin{equation}\label{finetunedK}
    K(\Phi, \bar{\Phi}, X, \bar{X})=K_{\text{main}}(\Phi, \bar{\Phi}) + \frac{M^2}{M^2 + V_{\text{inf}}(\Phi)} X \bar{X}
\end{equation}
and
\begin{equation}\label{SP1}
    W(\Phi, X)=W_{\text{main}}(\Phi) + W_0 + M X
\end{equation}
where $K_{\text{main}}$ and $W_{\text{main}}$ are subject to the following conditions that will ensure that the corresponding SUGRA model has a supersymmetric Minkowski vacua with flat directions. These conditions are as follows;
\begin{equation}\label{SIC1}
    K_{\text{main}}(\Phi, \bar{\Phi}) = 0, \;\;\; \partial_{\Phi} K_{\text{main}}(\Phi, \bar{\Phi}) = \partial_{\bar{\Phi}} K_{\text{main}}(\Phi, \bar{\Phi})= 0
\end{equation}
and
\begin{equation}\label{SIC2}
    W_{\text{main}}(\Phi) = 0, \;\;\; \partial_{\Phi} W_{\text{main}}(\Phi) = 0,
\end{equation}
at flat directions $\Phi - \bar{\Phi} = 0$. Here the partial derivatives are denoted as $\partial_{\Phi} \equiv \partial / \partial \Phi$, and it can be seen that there may be a shift symmetry on $\Phi$ for these conditions to be satisfied. For this model, the SUGRA potential at the inflationary trajectory appears as
\begin{equation}\label{SecondSUGRAPot}
    V(\Phi)=   V_{\text{inf}}(\Phi) +M^2 -3 W_{0}^{2}.
\end{equation}
\hspace{4mm} Different from the previous one (\ref{FirstSUGRAPot}), this potential holds an actual constant term for the cosmological constant, and also a separate term for an arbitrary scalar inflation potential. Therefore, $\Lambda =  M^2 -3 W_{0}^{2} $ being the cosmological constant, the scalar inflation potential can reach zero and disappear at its minimum, as it is for the most inflationary models. Note that, as long as these conditions are satisfied, any string-derived Kähler potential and superpotential would lead to same conclusion in terms of the SUGRA potential \cite{Kallosh:2021vcf}. Therefore, this model unlocks a new class of inflationary models that work in harmony with string theory. However, in case one excludes the $\overline{D3}$ brane contributions by taking $M \rightarrow 0$ limit, one obtains a supersymmetric AdS vacua in case scalar potential rolls to a zero vev or becomes smaller than $3 W_{0}^2$. Another issue with this model is naturalness since the bilinear nilpotent coupling of (\ref{finetunedK}) is fine-tuned to achieve the desired uplifting for a positive and small cosmological constant.

\subsection{Linear Nilpotent Corrections}\label{Sec2.2}
\hspace{4mm} One can denote the most general form of the Kähler potential with the main part that arises from string theory in addition to linear and bilinear nilpotent corrections as
\begin{equation}\label{MostGenKahler}
    K (\Phi, \bar{\Phi}, X, \bar{X})= K_{\text{main}}(\Phi, \bar{\Phi}) + f(\Phi,\Bar{\Phi}) X \Bar{X} + g(\Phi,\Bar{\Phi}) X + \bar{g}(\Phi,\Bar{\Phi}) \Bar{X}.
\end{equation}
Then, given the conditions (\ref{SIC1}) and (\ref{SIC2}) for its string derived main part, the most general form of the SUGRA potential (at the inflationary trajectory) becomes
\begin{equation}\label{MostmostGenKah}
    V=\frac{\left(W_{0} g + M\right) \left(W_{0} \bar{g} + M\right)}{f- \frac{\partial_{\Phi}\bar{g} \partial_{\bar{\Phi}}g}{\partial_{\Phi}\partial_{\bar{\Phi}} K_{\text{main}}}} - 3 W_{0}^{2}
\end{equation}
with the same superpotential (\ref{SP1}). Here we take $f \equiv f(\Phi,\bar{\Phi})\big|_{\Phi=\bar{\Phi}}$, $g \equiv g(\Phi,\bar{\Phi})\big|_{\Phi=\bar{\Phi}}$  with both linear and bilinear nilpotent Kähler correction terms included. In this potential, derivatives of linear Kähler corrections interact with the Kähler metric, $\partial_{\Phi}\partial_{\bar{\Phi}} K_{\text{main}}\big|_{\Phi=\bar{\Phi}}$ that created from the main part of the Kähler potential. The F-terms at flat directions can be denoted by
\begin{equation}\label{Fterms}
    \mathcal{D}_{\Phi} W =0 \;\;\; \text{and} \;\;\;  \mathcal{D}_{X} W = M + g(\Phi,\bar{\Phi}) W_{0},
\end{equation}
and the SUSY is spontaneously broken along the $X$-direction. As in the previous scenarios, considering only the bilinear corrections by taking $g=0$ and $f\neq 0$ at the beginning, reduces this SUGRA potential to
\begin{equation}\label{VsugraLessGeneral}
    V=\frac{M^2}{f} - 3 W_0^2
\end{equation}
where $f(\Phi)=1/V_{\text{inf}}(\Phi)$ leads to the first scenario and $f(\Phi)=M^2/(M^2 + V_{\text{inf}}(\Phi))$ to the latter. Note that in this case, F-term along the $X$-direction is fixed with $\mathcal{D}_{X} W = M$. Similarly, one can see that in case the function $g(\Phi,\bar{\Phi})$ is subject to the similar conditions at flat directions ($\Phi = \bar{\Phi}$) as
\begin{equation}
    g(\Phi, \bar{\Phi}) = 0, \;\;\; \partial_{\Phi} \bar{g}(\Phi, \bar{\Phi}) = \partial_{\bar{\Phi}} g(\Phi, \bar{\Phi})= 0,
\end{equation}
as the main part of the Kähler potential, the resulting SUGRA potential would be the same as (\ref{VsugraLessGeneral}). In this sense, there are a few possible branches that one can explore, with respect to the behaviour of the linear couplings. For simplicity, let us focus to the branch where $g(\Phi, \bar{\Phi}) \neq 0$ and $\partial_{\Phi}\bar{g}(\Phi, \bar{\Phi}) = \partial_{\bar{\Phi}} g(\Phi, \bar{\Phi})= 0$ at flat directions. For this branch, one can see that the derivatives $\partial_{\bar{\Phi}}g$ and $\partial_{\Phi}\bar{g}$ disappear from the SUGRA potential together with the term $\partial_{\Phi}\partial_{\bar{\Phi}} K_{\text{main}}$. Therefore the SUGRA potential becomes insensitive to whether the theory is canonical or not.
\vspace{-3mm}
\paragraph{Real spectrum:}
In case $g(\Phi,\bar{\Phi})=\bar{g}(\Phi,\bar{\Phi})$ (purely real), then it is coupled with the real part of the superfield $\Phi$. Here, the resulting potential becomes
\begin{equation}
    V=\frac{\left(M + W_0 g \right)^2}{f} -3 W_0^2 .
\end{equation}
At this point, one can generalize the SUGRA potential for a full spectrum of constant linear corrections with $g \propto \mu$ where $\mu$ is a real valued constant. An appropriate coupling choice as
\begin{equation}
    g=\frac{\mu - M}{W_0}
\end{equation}
would lead to the SUGRA potential 
\begin{equation}\label{SugraPotGen}
    V=\frac{\mu^2}{f} - 3 W_0^2.
\end{equation}
Here the coupling of the bilinear corrections can be generalized for both scenarios as
\begin{equation}\label{GenBilinearCor}
    f(\Phi)=\frac{\mu^2}{M^2 V_{\text{inf}}(\Phi)} \quad \text{or} \quad f(\Phi)=\frac{\mu^2}{M^2 + V_{\text{inf}}(\Phi)}
\end{equation}
with same order. As one can see, for $\mu \rightarrow M$ (or equivalently $g \rightarrow 0$) the coupling of the linear corrections disappear and one gets the previous formula (\ref{VsugraLessGeneral}) for the SUGRA potential. The value of $\mu$ also determines whether linear corrections are positive ($\mu > M$) or negative ($\mu < M$), although the resulting SUGRA potential is insensitive to the sign of the linear corrections. 
\vspace{-3mm}
\paragraph{Imaginary spectrum:}
So far, we have generalized the SUGRA potential for the purely real (constant) spectrum of linear couplings. The symmetries common to the cosmological models reviewed above are the shift symmetry $\Phi \rightarrow \Phi + a$, where $a$ has a real value and $\Phi \rightarrow \bar{\Phi}$ is the reality condition. To use the same method and generalize the SUGRA potential for the imaginary spectrum of linear couplings, firstly, we need to reveal an interesting case that will coincide with a symmetry condition imposed on the nilpotent superfield. Considering the linear terms for $X$ and $\bar{X}$ of the nilpotently corrected Kähler potential
\begin{equation}
     K (\Phi, \bar{\Phi}, X, \bar{X})= ... + g(\Phi,\Bar{\Phi}) X + \bar{g}(\Phi,\Bar{\Phi}) \Bar{X},
\end{equation}
one can make the choice for the arbitrary non-holomorphic function as $g(\Phi,\Bar{\Phi}) = -\bar{g}(\Phi,\bar{\Phi})$ for a purely imaginary coupling term. For this case, linear corrections are coupled to the imaginary part of $\Phi$. Then the Kähler potential gets the linear nilpotent corrections as
\begin{equation}
    K (\Phi, \bar{\Phi}, X, \bar{X})= ... +  g(\Phi,\Bar{\Phi}) \left(X - \bar{X} \right).
\end{equation}
Here one can see that, for a purely imaginary linear coupling term as $g(\Phi,\bar{\Phi})=\Phi-\bar{\Phi}$, in order to preserve reality condition $\Phi \rightarrow \bar{\Phi}$, one should also impose the symmetry $X \rightarrow -X$ (or $X \rightarrow \bar{X}$) on the nilpotent superfield \textit{simultaneously}. Moving on with the purely imaginary linear corrections, the general form of the potential becomes
\begin{equation}
    V=\frac{M^2 - W_0^2 g^2 }{f} -3 W_0^2.
\end{equation}
As like in the purely real case, one can chose a coupling term as $g\propto i \mu$ in order to obtain the full spectrum of constant and purely imaginary linear corrections. For example, in order to be consistent with the purely real case, one can denote that
\begin{equation}
        g=\pm i \frac{\sqrt{\mu^2 - M^2}}{W_0} \quad \text{for} \quad  |\mu| > |M|.
\end{equation}
Then, the generalized bilinear couplings given with (\ref{GenBilinearCor}) would lead to the same SUGRA potential with the purely real case $g(\Phi,\bar{\Phi})=\bar{g}(\Phi,\bar{\Phi})$. That means one can characterize the SUGRA potential with equation (\ref{SugraPotGen}) for all purely real ($g\propto \mu$) and imaginary ($g\propto i\mu$ when $|\mu|>|M|$) values of the constant linear couplings. Note here in particular for both examples of constant linear couplings, one equivalently obtains the same results via a suitable Kähler transformation as expected. Thus, in this general form, the constant parameter $\mu$ takes the place of the $\overline{D3}$ brane contribution $M$ and presents a SUGRA potential that is controlled by (constant) linear nilpotent corrections or an equivalent Kähler transformation.
\vspace{-3mm}
\paragraph{Real and Imaginary spectrum with scalars:}
At this point, we obtained SUGRA potentials for the constant linear corrections, and one can get the above models by an appropriate choice of the bilinear couplings to nilpotent superfields. However, one can still expect to see what happens when the linear couplings are $\Phi$ dependent, and therefore the linear corrections are not constant. Following that expectation, there is a possible case that fits in the same branch where the linear correction terms appear with a function of a single superfield such as $g\equiv g(\Phi)$ and $\bar{g}\equiv \bar{g}(\bar{\Phi})$. For this case, the SUGRA potential becomes
\begin{equation}\label{VGEN}
    V=\frac{\left(W_{0} g + M\right) \left(W_{0} \bar{g} + M\right)}{f} - 3 W_{0}^{2}.
\end{equation}
In case the linear terms appear with generic functions for the real or imaginary directions as
\begin{equation}\label{g1ng2}
    g(\Phi) =\frac{\mu h(\Phi) - M}{W_0} \quad \text{or} \quad g(\Phi) =\frac{i \mu h(\Phi) - M}{W_0},
\end{equation}
one obtains the same SUGRA potential as 
\begin{equation}\label{SugraGenPot2}
    V=\frac{\mu^2 h^2}{f} - 3 W_{0}^{2},
\end{equation}
for both linear couplings.\footnote{It is possible to see that given a linear coupling term mixed with an imaginary part as in (\ref{g1ng2}), one needs to impose a simultaneous $X\rightarrow \bar{X}$ symmetry on the nilpotent superfield to keep the Kähler potential invariant under a $\Phi \rightarrow \bar{\Phi}$ symmetry.} Here, we denoted the generic functions of the linear corrections as $h^2 \equiv h(\Phi) h(\bar{\Phi}) \big|_{\Phi=\bar{\Phi}}$. It can be seen that for different values of the constant coupling term $\mu$, all real and imaginary spectra of linear corrections are obtained by a generic function $h(\Phi)$. Thus, one can obtain various cosmological scenarios with a suitable combination of the bilinear and linear nilpotent couplings where both are functions of the inflation multiplet with $\Phi$. In this case, a generic inflation potential may also arise with the linear corrections alone. For example, an appropriate choice of the generic function with $h(\Phi)h(\bar{\Phi})=M^2 V_{\text{inf}}(\Phi,\bar{\Phi})$ would lead to the first scenario with the bilinear corrections as $f=\mu^2$, as a constant. However, since the F-term along the $X$-direction becomes
\begin{equation}
    \mathcal{D}_{X} W \big|_{\Phi=\bar{\Phi}} =  \mu h(\Phi)\big|_{\Phi=\bar{\Phi}} \quad \text{or} \quad \mathcal{D}_{X} W \big|_{\Phi=\bar{\Phi}} =  i \mu h(\Phi)\big|_{\Phi=\bar{\Phi}} ,
\end{equation}
SUSY would be restored if potential reaches a minimum at $h(\Phi)\big|_{\Phi=\bar{\Phi}} \rightarrow 0$, and one obtains a supersymmetric AdS vacuum at the minimum.

\vspace{3mm}
 
For all the SUGRA potentials described above, a general scalar potential should appear in the bilinear and (or) linear nilpotent Kähler corrections. But this appears unnatural since one needs to fine-tune these couplings beforehand to obtain the desired results, i.e. an almost vanishing cosmological constant with a dS vacua. It is reasonable, then, to ask $\bf i.$ if there is a way to obtain inflation and the desired uplifting more naturally with nilpotent Kähler corrections and $\bf ii.$ if there are any bounding conditions on these general nilpotent couplings. The answers are as follows:
\begin{itemize}
\item[\textbf{i.}] In string theory (or SUGRA), the corresponding field space metric may not always lead to canonical kinetic terms. Therefore, it makes more sense to consider the starting point with non-canonical fields to capture the broader context of the theory. With this in mind, we have produced a (super)universal attractor mechanism for inflation where all nilpotent coupling terms are obtained with a single parameter (in section \ref{SUAFNC}). Following that, we reveal a novel uplifting method by adding the same parameter as a complex contribution parallel to the decomposition of a superfield (in section \ref{Sec3.1}). Moreover, we analyzed a particular unified region of inflationary attractors (in section \ref{Sec3.2}), where dS vacua appear for (super)universal attractors without fine-tuning, leading to an almost vanishing cosmological constant.
\item[\textbf{ii.}] A direct consequence of these terms will appear on derivatives of both the Kähler and SUGRA potential. Thus, generic nilpotent coupling terms will also result in changes in the outcomes of the stability conditions and the mentioned dSC in a cosmological perspective for inflation. Therefore, we analyzed whether any bounding conditions existed to the generic nilpotent coupling terms of the (super)universal attractor mechanism considering stability conditions (in section \ref{Sec3.3}) and dSC (in section \ref{Sec3.4}). We show that for all values of the canonically normalized real scalar field, there is no bound on the only parameter of (super)universal attractor mechanism during the cosmological evolution. In addition, we show that the inclusion of nilpotent superfields to the dSC expanded the boundaries of the allowed EFT space.
\end{itemize}

\section{(Super)Universal Attractors}\label{SUAFNC}
\hspace{4mm}
Let us first briefly describe the universal attractor behavior in the (classical) inflation regime. It was first introduced in the Jordan frame with a novel non-minimal coupling to gravity \cite{Kallosh:2013tua},
\begin{equation}
    \mathcal{L}_{\text{J}} =\sqrt{-g} \left[\frac{R}{2} \Omega(\phi) - \frac{1}{2}\left(\partial \phi \right)^2 - \mu^2 h(\phi)^2\right].
\end{equation}
Here, $\Omega(\phi)=\left( 1+\xi h(\phi)\right)$ and $\phi$ is a real scalar field while $\xi$ and $\mu$ are real coupling constants. Obtaining the physical Lagrangian is easy with an appropriate conformal transformation as $g^{\mu \nu} \rightarrow \Omega(\phi) g^{\mu \nu} $ that leads to the Einstein frame with a non-canonical kinetic term as
\begin{equation}\label{NC_Classical_Lagrangian}
    \mathcal{L}_{\text{E}} =\sqrt{-g} \left[\frac{R}{2}  - \frac{3}{4} \left[\partial_{\phi} \log \Omega(\phi) \right]^{2} \left(\partial \phi \right)^2 - \frac{\mu^2 h(\phi)^2}{\Omega(\phi)^2}\right],
\end{equation}
at a strong coupling limit ($\xi \gg 1$). It is then possible with a field redefinition
\begin{equation}
    \left(\partial_{\phi} \varphi\right)^2 = \frac{3}{2} \left[\partial_{\phi} \log \Omega(\phi) \right]^{ 2}
\end{equation}
to map a diffeomorphism, and switch to the canonically normalized fields with
\begin{equation}\label{Diffeomorphism1}
   \varphi = \pm \sqrt{\frac{3}{2}} \log \Omega(\phi). 
\end{equation}
Then inserting these canonically normalized fields to the Lagrangian (\ref{NC_Classical_Lagrangian}) leads to an Einstein frame Lagrangian with canonical kinetic terms as
\begin{equation}\label{ClassicalAction}
     \mathcal{L}_{E} =\sqrt{-g} \left[\frac{R}{2}  - \frac{1}{2}  \left(\partial \varphi \right)^2 - \frac{  \mu^{2}}{\xi^{2}}  \left(1-\exp{\left(\mp \sqrt{\frac{2}{3}} \varphi\right)}\right)^{2}\right].
\end{equation}
 This model coincides with the scalar formulation of the $R+R^2$ Starobinsky model \cite{Starobinsky:1980te} (for minus sign) and results in the observational parameters of inflation as $n_{\text{s}}=1-2/N$ for the spectral index, and $r=12/N^2$ for the tensor-to-scalar ratio. Moreover, these parameters are in a highly consistent region of the latest observations \cite{Planck:2018jri,Aghanim:2018eyx,BICEP:2021xfz} for the e-foldings number $N \simeq 50\sim60$. 
 
On this account, these observational predictions are independent of the generic term $h(\phi)$; hence the diffeomorphism $\varphi \sim \log \Omega(\phi)$ and the non-canonical Einstein frame potential $V_{\text{E}}(\phi) \sim h\left(\phi\right)^2 / \Omega(\phi)^2$ plays a central role for the universal attractor behavior. Consequently, generic nilpotent Kähler corrections are natural candidates to capture this universal attractor mechanism. Considering higher dimensional theories such as string theory, which involves compactification of the extra dimensions collected on a compact manifold, the dS part of the theory can be represented as
\begin{equation}\label{EFL}
    \mathcal{L}_{\text{E}} =\sqrt{-g} \left[\frac{R}{2}  - K_{\Phi \bar{\Phi}} \partial_{\mu} \Phi \partial^{\mu} \bar{\Phi} - V\left(\Phi,\bar{\Phi}\right) \right]
\end{equation}
with a $4d$ SUGRA Lagrangian, which includes a non-canonical Kähler metric, $K_{\Phi \bar{\Phi}}\big|_{\Phi=\bar{\Phi}} \neq 1$. Then, one can embed the Einstein frame potential of universal attractor Lagrangian (\ref{NC_Classical_Lagrangian}) to $\mathcal{N}=1$, $D=4$ SUGRA by considering the scalar part of (\ref{SugraGenPot2}) as
\begin{equation}
    V_{\text{scalar}}(\phi)=\frac{\mu^2 h(\Phi)h(\bar{\Phi})}{f(\Phi, \bar{\Phi})}\bigg|_{\Phi=\bar{\Phi}}  \equiv \frac{\mu^2 h(\phi)^2}{\Omega(\phi)^2}.
\end{equation}
Here, we take the scalar field $\phi$ as the real part of the superfield $\Phi$. For this case, an obvious choice for the bilinear and linear couplings can be provided by $f(\Phi, \bar{\Phi})\big|_{\Phi=\bar{\Phi}} \equiv \Omega(\phi)^2$ and $h(\Phi)h(\bar{\Phi})\big|_{\Phi=\bar{\Phi}} \equiv h(\phi)^2$. That was the first step of embedding the universal attractor mechanism to nilpotent SUGRA. For the second step, it remains to capture the same non-canonical kinetic terms from SUGRA's universal attractor mechanism. Comparing with the (classical) inflation Lagrangian (\ref{NC_Classical_Lagrangian}) to map the same diffeomorphism on flat directions,
\begin{equation}
    K_{\Phi \bar{\Phi}} \partial_{\mu} \Phi \partial^{\mu} \bar{\Phi} \big|_{\Phi=\bar{\Phi}} =  \frac{1}{2} \left(\partial \varphi \right)^2,
\end{equation}
one needs to define a differential equation for
\begin{equation}\label{AttractorDif}
    K_{\Phi \bar{\Phi}}\big|_{\Phi=\bar{\Phi}} = \frac{3}{4} \left[\partial_{\phi} \log \Omega(\phi) \right]^{ 2}
\end{equation}
upon identifying the real part of the original multiplet as Re$(\Phi) = \phi$. Since the required solution is only for the (classical) inflationary Lagrangian in flat directions, this raises an uncertainty of possible Kähler potentials in its general form. Firstly, by defining the kinetic part of the Lagrangian as 
\begin{equation}
    K_{\Phi \bar{\Phi}} \partial_{\mu} \Phi \partial^{\mu} \bar{\Phi} =  \partial_{\mu} \Psi \partial^{\mu} \bar{\Psi},
\end{equation}
one must obtain solutions for the canonical field $\Psi$ by solving $K_{\Phi \bar{\Phi}} \equiv \partial_{\Phi} \Psi \partial_{\bar{\Phi}} \bar{\Psi}$, where we take the real part of the canonical field as Re$(\Psi) = \varphi / \sqrt{2}$. However, non-canonical superfields can be associated with canonical superfields provided by $\Psi \equiv \Psi \left(\Phi \right)$ or $\Psi \equiv \Psi \left(\Phi, \bar{\Phi} \right)$, in contrast to the classical case where real fields are associated with $\varphi \equiv \varphi \left( \phi\right)$. The latter definition for the canonical superfields is also often used to derive different multiplets in SUGRA. As we continue, we will define the main parameter of the (super)universal attractor mechanism as $\Omega(\Phi) = (1+\xi h(\Phi))$, in term of the non-canonical superfields. 
\begin{itemize}
    \item[\textbf{i.}]
    Assuming that $\Psi \equiv \Psi \left(\Phi \right)$ for the first case, one can establish a simple solution using the freedom of this ambiguity and by defining the canonical superfields as
\begin{equation}\label{firstSdiff}
   \partial_{\Phi} \Psi = \pm  \sqrt{\frac{3}{4}} \partial_{\Phi} \log \Omega( \Phi) \rightarrow \Psi = \pm \sqrt{\frac{3}{4}} \log \Omega( \Phi),
\end{equation}
where the Kähler metric relates to the canonical superfields as $K_{\Phi \bar{\Phi}} = \partial_{\Phi} \Psi \partial_{\bar{\Phi}} \bar{\Psi}$ and leads to the desired diffeomorphism on flat directions with (\ref{AttractorDif}). Then after solving for $K_{\Phi \bar{\Phi}}$, the final expression of the Kähler potential reads as
\begin{equation}\label{K1}
    K(\Phi,\bar{\Phi}) = \frac{3}{4} \log \Omega( \Phi) \log \Omega( \bar{\Phi}) + F(\Phi) + F(\bar{\Phi}),
\end{equation}
where $F(\Phi)$ and $F(\bar{\Phi})$ are arbitrary functions of their arguments. These arbitrary functions also coincide with a Kähler transformation that leaves the Kähler metric invariant, which will be necessary for the sequestered inflation mechanism. Considering the conditions for sequestered inflation mechanism,
\begin{equation}
    K_{\text{main}}(\Phi, \bar{\Phi})\big|_{\Phi=\bar{\Phi}} = 0 \;\;\; \text{and} \;\;\; \partial_{\Phi} K_{\text{main}}(\Phi, \bar{\Phi})\big|_{\Phi=\bar{\Phi}} = \partial_{\bar{\Phi}} K_{\text{main}}(\Phi, \bar{\Phi})\big|_{\Phi=\bar{\Phi}}= 0,
\end{equation}
one can adjust these arbitrary functions as
\begin{equation}
    F(\Phi) = -\frac{3}{8} \left[\log \Omega( \Phi)\right]^2.
\end{equation}
Then for (\ref{firstSdiff}), this choice leads to a shift symmetric Kähler potential with
\begin{equation}
    K(\Psi,\bar{\Psi}) = \Psi \bar{\Psi} - \frac{1}{2} \left( \Psi^2 + \bar{\Psi}^2\right) = -\frac{1}{2} \left( \Psi - \bar{\Psi}\right)^2
\end{equation}
and becomes compatible for the sequestered inflation mechanism in the canonical form too. The selection for arbitrary functions should always be equivalent to a Kähler transformation as $K \rightarrow K + F + \bar{F}$. Thus we can employ it to the string-derived main part of the Kähler potential. 
    \item[\textbf{ii.}]
    Moving on for the case $\Psi \equiv \Psi \left(\Phi, \bar{\Phi} \right)$, a suitable field redefinition can be given with
\begin{equation}\label{secondSdiff}
  K_{\Phi \bar{\Phi}} \equiv \partial_{\Phi} \Psi \partial_{\bar{\Phi}}\bar{\Psi} =\frac{ 3\; \partial_{\Phi}\Omega(\Phi) \; \partial_{\bar{\Phi}}\Omega(\bar{\Phi})}{\left(\Omega(\Phi) +  \Omega(\bar{\Phi})\right)^2} \rightarrow K_{\Phi \bar{\Phi}}\big|_{\Phi=\bar{\Phi}} = \frac{3}{4} \left[\partial_{\phi} \log \Omega(\phi) \right]^{ 2},
\end{equation}
and the corresponding Kähler potential can be denoted as
\begin{equation}
    K(\Phi,\bar{\Phi}) = -\frac{3}{2} \log \left[\left(\frac{\Omega(\Phi) +  \Omega(\bar{\Phi})}{2} \right)^2\right] + F(\Phi) + F(\bar{\Phi}).
\end{equation}
Since for this case, the first part of the Kähler potential logarithmically depends on both $\Phi$ and $\bar{\Phi}$, the Kähler transformation needs to appear with logarithms as
\begin{equation}
    F(\Phi)=\frac{3}{2}\log \left[\Omega(\Phi)  \right],
\end{equation}
to equalize the contribution of the first part. Then the final form of the Kähler potential becomes
\begin{equation}\label{K2}
    K(\Phi,\bar{\Phi}) = -\frac{3}{2} \log \left[\frac{\left(\Omega(\Phi) +  \Omega(\bar{\Phi}) \right)^2}{4\Omega(\Phi) \Omega(\bar{\Phi})}\right] \rightarrow K(\Phi,\bar{\Phi}) \big|_{\Phi=\bar{\Phi}} = \log 1 = 0
\end{equation}
with non-canonical fields and fulfills the sequestered inflation conditions for the Kähler potential.
\end{itemize}
The main part of the Kähler potential is only responsible for producing the expected kinetic term for the universal attractor mechanism. It does not affect the non-canonical scalar potential for sequestered inflation. Consequently, various Kähler potentials could produce the same kinetic term in flat directions without disordering the scalar potential up to a Kähler transformation.
\subsection{Uplifted Potential and dS Vacua}\label{Sec3.1}
\hspace{4mm}
Now that we obtained the universal attractor mechanism, we can revisit the nilpotent part of the Kähler potential. Up to this point, we have only considered the scalar part of the SUGRA potential to capture the universal attractor mechanism through the nilpotent part of the Kähler potential. However, in the total SUGRA potential (\ref{SugraGenPot2}), there remains an additional $-3W_{0}^2$ term which would lead to an AdS vacuum at the minimum of the scalar potential. Nevertheless, one can assemble an uplifting mechanism (or an uplifted potential) via linear coupling terms of the nilpotent Kähler corrections and their connection to the universal attractor mechanism.

Given a constant value of $\mu = - \xi M$, linear and bilinear nilpotent couplings can be defined in the same way as the main component
\begin{equation}\label{UAmaincomp}
    \Omega(\Phi) = (1+\xi h(\Phi)) \rightarrow \Omega(\Phi)\big|_{\Phi=\bar{\Phi}}  =(1+\xi h(\phi))=\rm{e}^{\pm \sqrt{\frac{2}{3}} \varphi}
\end{equation}
of the universal attractor mechanism, up to a constant multiplier. That is also equivalent to the conformal transformation function in flat directions. Accordingly, the nilpotent couplings become
\begin{equation}
    g(\Phi) =\frac{\mu h(\Phi) - M}{W_0} \rightarrow g(\Phi) = -\sqrt{3 \alpha}  \Omega(\Phi), \;\;\; f(\Phi,\bar{\Phi}) = \Omega(\Phi) \bar{\Omega}(\bar{\Phi}),
\end{equation}
and hence all nilpotent corrections have the same form. To avoid any assumptions here, we set up the relation
\begin{equation}\label{MW0relation}
    \left|\frac{M}{\sqrt{3} W_0} \right|= \sqrt{\alpha},
\end{equation}
where the real constant parameter $\sqrt{\alpha}$ represents the ratio of $\overline{D3}$ brane contribution with $M$ and the gravitino mass with $W_0$. That also results in fixing the coupling constant of the Einstein frame potential as $M^2$, and inflation gets an explicit dependence on the $\overline{D3}$ brane contribution. At this point, the total SUGRA potential becomes:
\begin{equation}
         V\left(\Phi, \bar{\Phi}\right)\big|_{\Phi=\bar{\Phi}}  = M^2 \left(\frac{1}{\Omega(\phi)} -1 \right)^2 -\frac{M^2}{\alpha}.
\end{equation}
Then, the only necessary step to obtain an uplifted dS potential is to provide an explicit dependence on linear couplings with an imaginary part parallel to a decomposition of a superfield as;
\begin{equation}
    g(\Phi)=-\sqrt{3\alpha} \Omega(\Phi) \rightarrow g(\Phi)= -\sqrt{3\alpha} \left(\Omega(\Phi) + i \Omega(\Phi) \right).
\end{equation}
In terms of our new linear couplings, the nilpotent Kähler corrections become
\begin{equation}\label{Knc}
    \begin{aligned}
        K_{N.C.}= \Omega \bar{\Omega} X \bar{X} -\sqrt{3 \alpha} \left(\Omega + i \Omega\right) X -\sqrt{3 \alpha} \left(\bar{\Omega} - i \bar{\Omega}\right) \bar{X} 
                = \frac{1}{6\alpha} g \bar{g} X \bar{X} + g X + \bar{g} \bar{X}.
    \end{aligned}
\end{equation}
Thus, one obtains the total SUGRA potential with non-canonical fields as
\begin{equation}\label{NonCanPot1}
         V\left(\Phi, \bar{\Phi}\right)\big|_{\Phi=\bar{\Phi}}  = M^2 \left(\frac{1}{\Omega(\phi)} -1 \right)^2 + M^2\left(1-\frac{1}{\alpha}\right),
\end{equation}
where the exact uplifting amount is $M^{2}$, provided with the contribution of the term $i \Omega(\Phi)$ in the linear couplings. Here, one can see that the first term expresses the $R+R^2$ Starobinsky potential with non-canonical fields, and
\begin{equation}
    \Lambda = M^2\left(1-\frac{1}{\alpha}\right) = M^2 - 3 W_{0}^2
\end{equation}
is a constant residual term for the cosmological constant. That term results in a dS vacua at the minimum of the scalar potential with $\alpha > 1$ or equivalently with $\Lambda_{dS}=M^2 - 3 W_{0}^2 > 0$.
\begin{figure}[htb]
\centering
\includegraphics[width=3.8in]{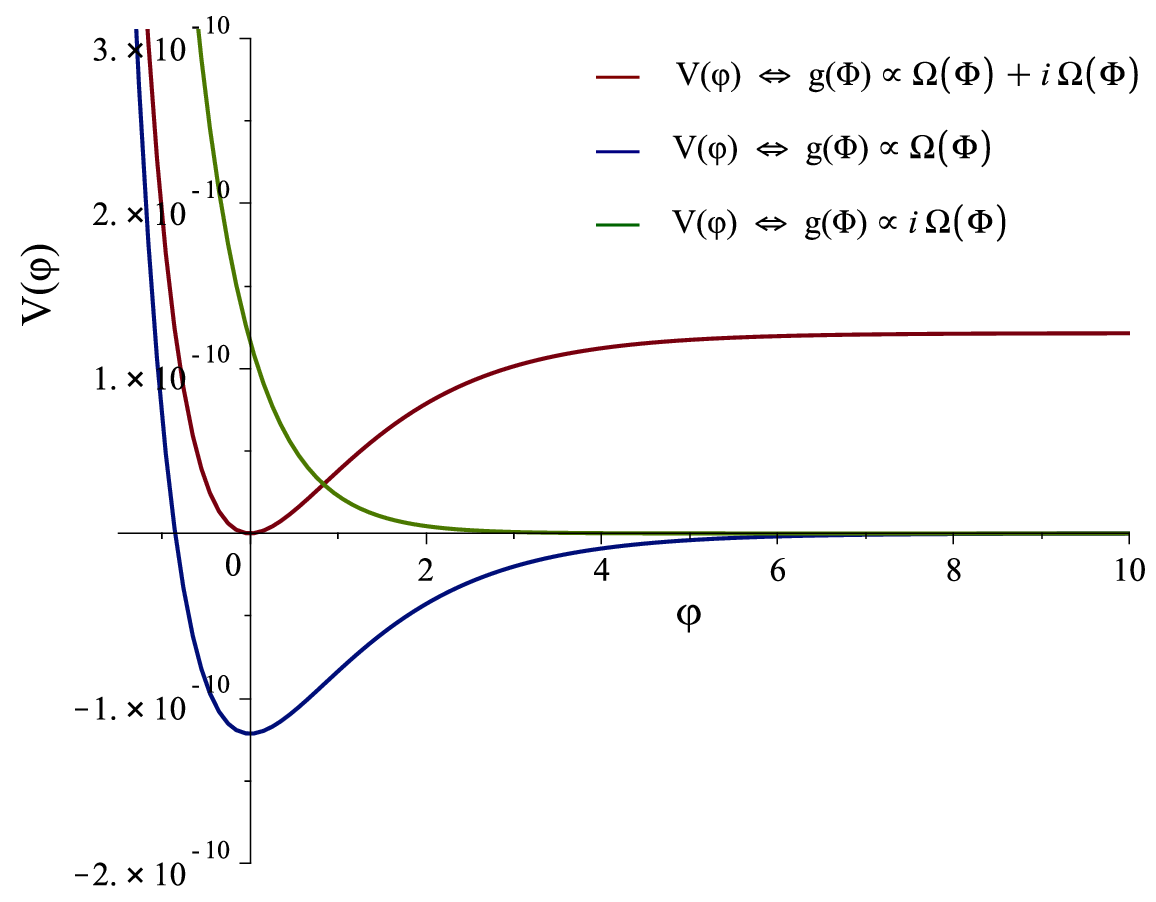}
\caption{\label{F1} 
Evolution of the F-term SUGRA potential on flat directions for different choices of linear nilpotent couplings. The bilinear nilpotent coupling term was established as $f(\Phi,\bar{\Phi}) = \Omega(\Phi) \Omega(\bar{\Phi})$ with the main component of the (super)universal attractor mechanism $\Omega(\Phi)\big|_{\Phi=\bar{\Phi}}  =\exp\left( \sqrt{\frac{2}{3}} \varphi\right)$. The blue line shows the potential before the uplift, and the green line solely illustrates the potential's uplift contribution. The red line shows the uplifted potential that leads to a dS minimum  with $M/\sqrt{3} \simeq W_0 \simeq \pm 6.36 \times 10^{-6}$ for an almost vanishing cosmological constant as $ \Lambda_{dS} =M^2 -3 W_{0}^2 \gtrsim 0 $. 
}
\end{figure}
Moving on, one can use (\ref{firstSdiff}) for a specific canonical superfield choice, or more generally, one can use the conformal transformation function (\ref{UAmaincomp}) in flat directions and switch to the real canonical fields as
\begin{equation}\label{CanPot}
     V(\varphi) =  M^2 \left(1 - \rm{e}^{\mp \sqrt{\frac{2}{3}} \varphi} \right)^2 + \Lambda_{dS}.
\end{equation}
Consequently, the $\overline{D3}$ brane contribution determines the coupling constant of the scalar potential. That being so, it is a common standpoint in inflationary cosmology to consider the quantum fluctuations of the inflaton field as responsible for the origin of the early universe. Hence, the observed value of the power spectrum of curvature amplitude (squared)
\begin{equation}
    \Delta_{\mathcal{R}}^2 = \frac{1}{12 \pi^2} \frac{V^3}{\left(\partial_{\varphi}V\right)^2}
\end{equation}
allows one to determine the coupling constant of $R+R^2$ Starobinsky potential as
\begin{equation}
    M^2 \simeq \frac{9 \pi^2 \Delta_{\mathcal{R}}^2}{8} \left( \frac{4N + 3}{N^2}\right)^2.
\end{equation}
Here, $\Delta_{\mathcal{R}}^2 \simeq 2.1 \times 10^{-9}$ from the Planck measurement \cite{Planck:2018jri,Aghanim:2018eyx} for the pivot scale $k_{*} = 0.05$ Mpc$^{-1}$, and $N$ is the number of e-folds. Thus, the quantum fluctuations of the inflaton determine the $\overline{D3}$ brane contribution for $N\simeq56.2$ as $M^2 \simeq 1.21 \times 10^{-10}$, prior to estimating an equally small gravitino mass (squared) $W_{0}^2 \simeq 4.04 \times 10^{-11}$ for a nearly vanishing cosmological constant with $M/\sqrt{3} \simeq W_0 \simeq \pm 6.36 \times 10^{-6}$.
\vspace{-3mm}
\paragraph{Avoiding supersymmetric AdS:}
Taking the Kähler potential (\ref{Knc}) and the superpotential (\ref{SP1}) as a starting point, one obtains a dS vacua ($\alpha>1$) and spontaneously broken SUSY without generating any supersymmetric AdS vacua. To see this more clearly, one can extract the $\overline{D3}$ brane contribution from the superpotential by taking the $M\rightarrow 0$ limit. 
\begin{itemize}
    \item[\textbf{i.}] If one considers the ratio (\ref{MW0relation}) for $\alpha$, thus interpreting the gravitino mass and $\overline{D3}$ brane contribution in a relation, the F-term along the $X$-direction becomes
\begin{equation}
    \mathcal{D}_{X} W = M-(i+1)\Omega M
    \rightarrow
    \lim_{M \to 0} \mathcal{D}_{X} W  = 0.
\end{equation}
Following that, one obtains a pure AdS state with restored SUSY, since the potential becomes
\begin{equation}
     V =  \left(\frac{M}{\Omega} -\left|\frac{M}{ W_0} \right| W_0 \right)^2 + M^2 - 3 W_{0}^{2} \rightarrow \lim_{M \to 0} V = -3 W_{0}^{2},
\end{equation}
without the $\overline{D3}$ brane contributions in the superpotential. 
 \item[\textbf{ii.}] If one interprets $\alpha$ as an independent parameter, the F-term along the same direction reads 
\begin{equation}
    \mathcal{D}_{X} W = M-(i+1)\Omega \sqrt{3 \alpha} W_{0}
    \rightarrow
    \lim_{M \to 0} \mathcal{D}_{X} W  =- (i+1)\Omega \sqrt{3 \alpha}W_{0}.
\end{equation}
Accordingly, the SUSY remains spontaneously broken at all times with $\Omega \neq 0$. Thus, the potential becomes
\begin{equation}
    V =  \left(\frac{M}{\Omega} -\sqrt{3\alpha} W_{0} \right)^2 + 3 W_{0}^{2}\left(\alpha - 1\right) \rightarrow \lim_{M \to 0} V = 6 W_{0}^2 \left(\alpha - \frac{1}{2} \right),
\end{equation}
leading to a pure dS or Minkowski state for $\alpha \geq 1/2$. Even though it results in a pure AdS state for $\alpha < 1/2$, SUSY remains spontaneously broken, and one avoids producing a supersymmetric AdS vacua.
\end{itemize}
Note that excluding the $\overline{D3}$ brane contribution $M$ for both interpretations removes the scalar part of the potential; thus, one can see that inflationary dynamics are only produced in the presence of $\overline{D3}$ branes. 

\subsection{Unification of Inflationary Attractors}\label{Sec3.2}
\hspace{4mm}
There is a profound connection between inflationary attractor models such as conformal attractors \cite{Kallosh:2013hoa,Kallosh:2013daa}, $\alpha$-attractors \cite{Ferrara:2013rsa,Kallosh:2013yoa}, and $\xi$-attractors \cite{Kallosh:2013tua,Bezrukov:2007ep,Giudice:2014toa,Pallis:2013yda,Pallis:2014dma,Pallis:2014boa}. One can connect these inflationary attractors in a unified region where the parameters of these attractors are associated together as $\alpha(\xi)$ and the inflationary predictions of all these models coincide in the large $\xi$ limit \cite{Galante:2014ifa}.

In that regard, one can move the parameter $\alpha$ in (\ref{Knc}) as a constant multiplier to all nilpotent Kähler corrections by rescaling the nilpotent superfields as $X \rightarrow \sqrt{\alpha} X$. With this rescaling the new bilinear and linear nilpotent couplings become
\begin{equation}
    f(\Phi, \bar{\Phi}) = \alpha \Omega(\Phi) \bar{\Omega}(\bar{\Phi}), \; \; \; g(\Phi) = -\sqrt{3} \alpha (\Omega(\Phi) +i \Omega(\Phi))
\end{equation}
giving rise to the Kähler potential
\begin{equation}\label{KncAlpha}
    \begin{aligned}
        K_{N.C.}= \alpha \left[\Omega \bar{\Omega} X \bar{X} -\sqrt{3 } \left(\Omega + i \Omega\right) X -\sqrt{3} \left(\bar{\Omega} - i \bar{\Omega}\right) \bar{X} \right]
                = \frac{1}{6\alpha} g \bar{g} X \bar{X} + g X + \bar{g} \bar{X},
    \end{aligned}
\end{equation}
without changing the structure of the Kähler potential in terms of the linear nilpotent couplings. Then, one can approach that part together with the main part of the Kähler potential and parameterize the total Kähler potential with the same parameter:
\begin{equation}\label{Knc2}
    K^{\prime} = \alpha \left[K_{\text{main}} + \Omega \bar{\Omega} X \bar{X} -\sqrt{3 } \left(\Omega + i \Omega\right) X -\sqrt{3} \left(\bar{\Omega} - i \bar{\Omega}\right) \bar{X} \right].
\end{equation}
Consequently, that allows us to capture an equivalence with the $\alpha$-attractor models since multiplying the main part with $\alpha$ for either (\ref{K1}) or (\ref{K2}) leads to the canonical transformation function 
\begin{equation}\label{ConformalTransformation}
    \Omega(\Phi)\big|_{\Phi=\bar{\Phi}}  =\rm{e}^{\pm \sqrt{\frac{2}{3 \alpha}} \varphi}
\end{equation}
and the SUGRA potential
\begin{equation}\label{CanPotAlpha}
     V(\varphi) =  \frac{M^{2}}{\alpha} \left(1 - \rm{e}^{\mp \sqrt{\frac{2}{3\alpha}} \varphi} \right)^2 + M^2\left(1-\frac{1}{\alpha}\right).
\end{equation}
The resulting spectral index shares an identical prediction with the universal attractors for the $\alpha$-attractor model. However, the tensor-to-scalar ratio becomes $r=12\alpha / N^2$ and sets a lower bound as $\alpha>0$. That is consistent with the relation (\ref{MW0relation}), between the $\overline{D3}$ brane contribution $M$ and the gravitino mass $W_0$, since $\sqrt{\alpha}$ is required to be a real constant parameter. 

Following this prospect, one can promote $\alpha$ to unified region of attractors where the inflationary predictions remains robust due to the properties of the leading pole in the Laurent expansion of the Einstein frame kinetic term. Thus, treating the conformal transformation function $\Omega(\phi)$ as a field variable\footnote{That approach is analogous to the uplifting mechanism where we decomposed $\Omega(\Phi)$ into real and complex parts similar to a superfield.} with a well-chosen Jordan frame kinetic term, one can denote \cite{Galante:2014ifa}:
\begin{equation}\label{UniLim}
    \alpha = 1 + \frac{1}{6 \xi},
\end{equation}
 with the condition $\xi > 0$ to obtain a positively signed kinetic term in the Jordan frame. This class of $\alpha$-attractors is fully equivalent to the $\xi$-attractors, and results in universal attractor model when $\xi \rightarrow \infty$, as also used in the derivation of the Lagrangian (\ref{NC_Classical_Lagrangian}) in the process of constructing (super)universal attractors. That also indicates that in the large $\xi$ limit where all the attractors unify, one fulfills $\alpha \gtrsim 1$ and obtains an almost vanishing cosmological constant ($\Lambda_{dS} = M^2 - 3 W_{0}^2 \gtrsim 0$) as in our universe.

\subsection{Moduli Stabilization}\label{Sec3.3}
\hspace{4mm}
To achieve a successful single field inflation period, one needs to consider stabilizing moduli fields with the correct hierarchical order. For example, for a standard single-field slow-roll inflation where the fluctuations of the inflaton field are produced with a light inflaton mass, one must verify that all other fields are heavier than the Hubble mass. With this in mind, the second derivatives of the potential (\ref{NonCanPot1}) can be given in terms of non-canonical fields ($\Phi = \phi + i \; \text{Im} \Phi$) as 
\begin{equation}
    \partial^{2}_{\phi} V\big|_{\Phi=\bar{\Phi}} = \frac{2 M^2}{\Omega^4} \left[    \left( \partial_{\phi} \Omega \right)^2 \left(3-2\Omega \right) + \left( \partial_{\phi}^2 \Omega \right) \left( \Omega^2 - \Omega\right)   \right]
\end{equation}
and
\begin{equation}
    \partial^{2}_{\text{Im} \Phi} V\big|_{\Phi=\bar{\Phi}} \simeq \frac{W_{0}^{2}}{\Omega^4} \left[    \left( \partial_{\phi} \Omega \right)^2 \left(5 \Omega^2 -2\Omega +1 \right) -2 \left( \partial_{\phi}^2 \Omega \right) \left( \Omega^2 - \Omega\right)   \right]
\end{equation}
 assuming that $M \simeq \sqrt{3} W_{0} $ for a nearly vanishing cosmological constant. Then, one can derive the mass terms of the canonical fields with
\begin{equation}
    \sqrt{2} \partial_{\phi} \equiv \partial_{\varphi} \;\;\; \text{and} \;\;\; \Omega(\phi) = \rm{e}^{\mp \sqrt{\frac{2}{3}} \varphi}.
\end{equation}
Equivalently, one can decompose the superfield $\Psi$ in terms of the real canonically normalized fields as
\begin{equation}
    \Psi = \frac{1}{\sqrt{2}} \left( \varphi + i \beta\right),
\end{equation}
and consider the SUGRA potential (\ref{NonCanPot1}) with canonical fields (\ref{firstSdiff}). 
\begin{figure}[htb]
\centering
\includegraphics[width=5in]{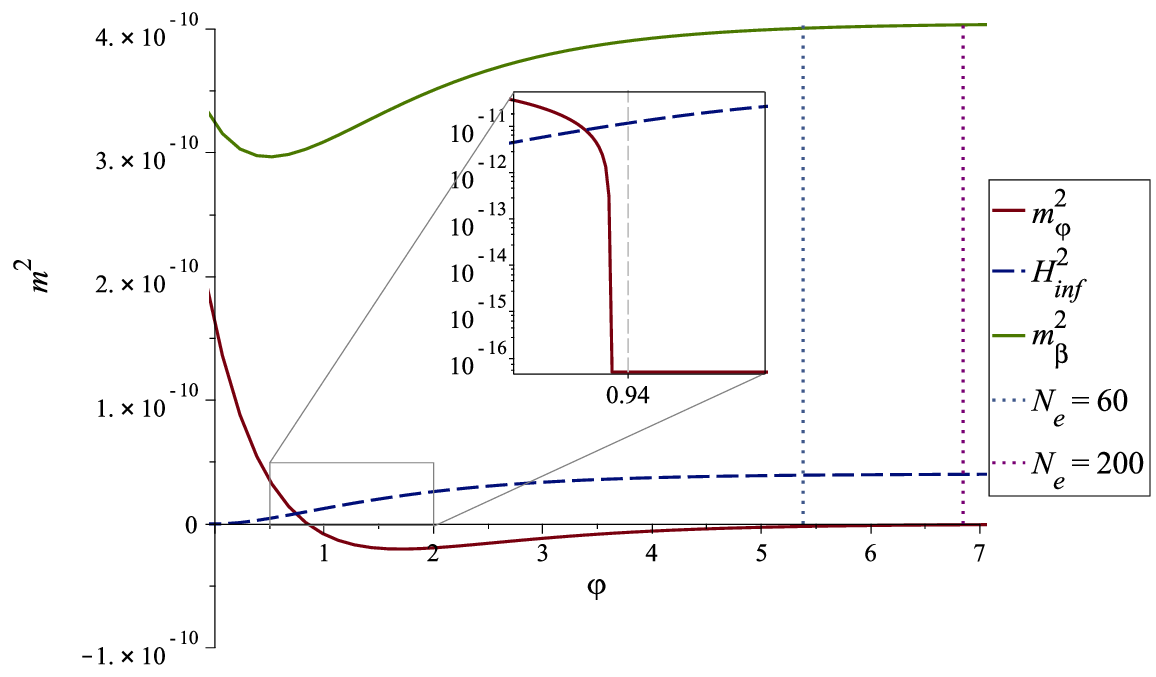}
\caption{\label{F2} The mass values for canonically normalized real fields during the cosmological evolution. The zoomed-in portion of the figure shows that the (absolute value of) inflaton mass (squared) $m_{\varphi}^2$ stays lighter than the Hubble mass during inflation which $\epsilon(\varphi_{\text{end}}=0.94) =1$ corresponds to the end of inflation. The imaginary portion of the mass is denoted by $m_{\beta}^2$ and always remains heavier than the Hubble mass ($H_{\text{inf}}^2 \simeq V/3$). $N_{e}$ denotes the number of e-folds obtained during the slow-roll inflation.
}
\end{figure}
Following that, the mass terms (second derivatives of the potential) can be denoted in terms of the slow-roll inflation parameters as
\begin{equation}
    m_{\varphi}^2 = \eta V \;\;\; \text{and} \;\;\; m_{\beta}^2 \simeq W_{0}^2 \left(10 - 2 \rm{e}^{- \sqrt{\frac{2}{3}} \varphi}\right) - \sqrt{3 \epsilon} V
\end{equation}
where the slow-roll parameters are
\begin{equation}
    \epsilon \simeq \frac{1}{2} \left( \frac{\partial_{\varphi} V}{V}\right)^2, \;\;\; \eta \simeq \frac{\partial_{\varphi}^2 V}{V}
\end{equation}
and $V$ denotes the $R+R^2$ Starobinsky potential for $\Omega(\phi) = \rm{e}^{ \sqrt{\frac{2}{3}} \varphi}$. Accordingly, choosing the real part of the superfield as the inflaton field, one needs to fulfill the conditions 
\begin{equation}\label{stability}
    m_{\varphi}^2 < H^2 \;\;\; \text{and} \;\;\; H^2 \lesssim m_{\beta}^2
\end{equation}
to obtain the correct hierarchical order of masses. Since both mass terms depend on the canonically normalized field $\varphi$, it is necessary to consider the entire evolution of these mass terms. As shown in figure (\ref{F2}), the absolute value of the inflaton mass remains lighter than the Hubble mass. (The inflaton mass is expected to take negative values during inflation since the observational constraints from the $n_s - r$ contours rule out the convex potentials leading to $\eta>0$.) The mass of $\beta$ always remains heavier than the Hubble mass, and consequently, the stability conditions (\ref{stability}) are satisfied. After the inflation, all mass values evolve to positive values and become stable at the dS minimum of the potential.

\subsection{String Landscape and Swampland}\label{Sec3.4}
\hspace{4mm}
The absence of the dS vacua in string theory led to the proposal of the dSC \cite{Obied:2018sgi,Garg:2018reu,Ooguri:2018wrx};
\begin{equation}
    |\nabla V| \geq c V \;\;\; \text{where} \;\;\; c \sim \mathcal{O}(1).
\end{equation}
That conjecture implies that for an extremum of a supersymmetric potential, one can not have a positive cosmological constant;
\begin{equation}
    |\nabla V| = 0 \rightarrow V \leq 0.
\end{equation}
The set of low-energy effective field theories consistent with this conjecture is thought to be in the \textit{Landscape}, otherwise in the \textit{Swampland}. One can rearrange this inequality as
\begin{equation}
    \epsilon_{V} = \frac{1}{2}\left( \frac{\nabla V}{V}\right)^2 \geq \frac{c^2}{2}.
\end{equation}
At this point, it is useful to make a clear distinction by naming the Hubble slow-roll parameter with $\epsilon \equiv \epsilon_{H}$, and the potential slow-roll parameter with $\epsilon_V$ as defined and investigated in \cite{Liddle:1994dx}. The Hubble slow-roll parameter is the precise physical condition for inflation and is required to be smaller than unity as $\epsilon_{H}<1$. On the other hand, the potential slow-roll parameter is proportional to the gradient flow of the potential. For a standard slow-roll inflation model evaluated on the Friedmann-Lemaître-Robertson-Walker (FLRW) background with a single scalar field that depend only on time, the Hubble slow-roll parameter becomes approximately equal to the potential slow-roll parameter,
\begin{equation}
    \epsilon \equiv \epsilon_{H} = \frac{\dot{\varphi}^2}{2 H^2} \simeq \frac{1}{2} \left( \frac{\partial_{\varphi} V}{V}\right)^2 = \epsilon_{V}.
\end{equation}
As a result, the condition for single-field slow-roll inflation, $\epsilon_{H} \ll 1 $, and the landscape condition for dSC, $\epsilon_{V} \geq c^{2}/2$, can not be satisfied simultaneously. (See appendix \ref{A.1} for a general multi-field case.)

Nevertheless, the Hubble slow-roll parameter is independent of the approximations used above, and it states that a decreasing Hubble radius is needed for an inflationary epoch in the early universe. Therefore, the Hubble slow-roll parameter should be considered for slow-roll criteria, as it is the correct description of inflation. Accordingly, depending on multi-field behavior, the approximate equivalence $\epsilon_{H} \simeq \epsilon_{V}$ becomes invalid in the broader picture for the possible multi-field inflationary scenarios, as also pointed out and discussed in the swampland context in \cite{Achucarro:2018vey}. Thus, theories with multi-field origins open up the possibility of simultaneously obtaining
\begin{equation}
    \epsilon_{H} \ll 1 \;\;\; \text{and} \;\;\; \epsilon_{V} \geq  \mathcal{O}(1)
\end{equation}
without causing a contradiction. Given this motivation, nilpotent SUGRA offers a possibility of overcoming this problem, as its structure suggests, by including nilpotent superfields in the potential gradient flow. 

In this section, we will analyze this possibility, considering the following; after deriving the SUGRA action, the initial approach to consider nilpotent superfields to derive the scalar potential is to remove them from the theory by hand. This can be justified as the nilpotent superfields have no scalar field contribution or vev. As a result, one can obtain an effective single-field inflation model evaluated on a flat direction as the inflationary trajectory at $X=0$. Equivalently, it is also appropriate to consider a unitary gauge imposed on goldstino fermions to remove nilpotent superfields. This gauge fixing justifies the former approach, but moreover, it opens the possibility of extending the validity of constrained superfields to cosmological regimes. Fortunately, it is possible to keep nilpotent superfields even after deriving the SUGRA action since physical equations of motion for scalar fields remain equivalent with or without nilpotent superfields. Thus, the potential gradient flow is allowed to be evaluated as a multi-field theory, including nilpotent superfields. After that, the same unitary gauge can be imposed to extract nilpotent superfields and see the predictions of this scalar theory. Accordingly, we will show the following two possible scenarios; 
\begin{itemize}
    \item[\textbf{i.}] The initial approach presents a viable single-field inflation model by extracting nilpotent superfields manually or applying a unitary gauge right after obtaining the action. Yet, the model falls into the swampland as expected.
\item[\textbf{ii.}] The latter provides a multi-field theory for the potential gradient flow and leads to the landscape for dSC upon applying the unitary gauge after taking derivatives. However, the viability of inflation depends on the imposition of $\partial_{t} X \equiv \dot{X}=0$, considering there are no time-dependent scalar contributions of nilpotent superfields.
\end{itemize}
For that matter, let us briefly summarise some of the technical details regarding the application of constrained superfields in cosmology, as also applied and described in the references given in the introduction and throughout the study.

\vspace{-3mm}
\paragraph{Nilpotent Superfields:}
In the context of cosmology with nilpotent superfields \cite{Ferrara:2014kva,Ferrara:2015tyn}, nilpotent SUGRA is an extension of the original Volkov-Akulov goldstino theory. The scalar part of superfield $X$ is given with
\begin{equation}
    X(\chi,F)= \frac{\bar{\chi} P_{L}\chi}{2F}\equiv \frac{\chi^2}{2F},
\end{equation} 
where the supersymmetric partner of the goldstino fermion (sgoldstino) within the original theory is replaced with a bilinear combination of goldstino fermions. The auxiliary field ($F$-term) is required to be non-vanishing for the $X$ direction where the SUSY is always broken. Since a bilinear combination of fermions replaces the scalar part of $X$, \textit{the field itself does not disappear}, but its expectation value vanishes \cite{Kallosh:2014via}. Unlike the conventional multi-field cases, the stability of the nilpotent superfield becomes irrelevant (i.e., it is not needed to stabilize at the minimum of its potential and then integrated outside of the spectrum), and one does not need to investigate the time evolution of that field for the description of inflation on the bosonic part of the theory, i.e., $\dot{X} = 0$. 

More explicitly, this can be ensured by imposing that $X=0$ for removing the non-scalar nilpotent superfields from the scalar theory after obtaining the SUGRA action. This is also consistent with the nilpotency constraint $X^2=0$, which appears as an equation of motion for nilpotent SUGRA, indicating that the product of the Grassmann variables of the fermionic bilinear must disappear. Consequently, $X$ does not have any vev. After this imposition, the stabilized Kähler moduli fields can be treated on the inflationary trajectory at $X=0$ without considering the stability of constrained superfields.

\vspace{-3mm}
\paragraph{Inflationary Trajectory:}
In particular, a variety of effective single-field inflation models can be realized from multi-field theories by choosing a \textit{stable trajectory} and killing out all the scalar fields but one. Thus, the (independent) dynamical evolution of all the other scalar fields is excluded from equations of motion for the corresponding effective single-field model. For example, one can decompose the superfield $\Psi$ in terms of real canonically normalized scalar fields as 
\begin{equation}
    \Psi \propto \left(\varphi + i \beta \right).
\end{equation}
Then, one can evaluate the theory on flat directions with $\Psi - \bar{\Psi}=0$, corresponding to a trajectory for real fields as $\varphi  \neq 0$, $\beta =0$. Here for the effective single-field model, $\varphi$ is chosen as the inflaton field, and the real scalar field $\beta$ must be stabilized on that trajectory and remain heavier than the Hubble mass at all times (as shown in Figure \ref{F2}). Note that, unlike the nilpotent superfields, $\beta$ is a real scalar field with a non-zero expectation value that evolves with cosmic time; hence, it must be stabilized. So far, that is the general approach in the literature, and consequently, nilpotent superfields are included in the inflationary trajectory as
 \begin{equation}
     X= \Psi - \bar{\Psi} =0.
 \end{equation}
With this inclusion, a swift transition can be made from the already existing multi-field originated models by simply declaring one of the superfields nilpotent. Then, in addition to the further distinctions as examined in this study, one can utilize the properties of nilpotent superfield $X$, such as being stabilization free since it does not acquire any vev and is set apart from the cosmological evolution.\footnote{One can also treat these fields as auxiliary Stueckelberg-like fields \cite{Ruegg:2003ps}; see also \cite{Boulanger:2018dau,Lyakhovich:2021lzy,Abakumova:2021evc} for some recent discussions.}

For this (initial) approach, given the scalar potential (\ref{CanPot}) at the inflationary trajectory, dSC can be evaluated for a single-field theory as the scalar formulation of the $R+R^2$ Starobinsky model since there will be no scalar fields associated with nilpotent superfields remaining in the potential. Hence one obtains \cite{Kehagias:2018uem,Dias:2018ngv,Matsui:2018bsy,Ben-Dayan:2018mhe,Kinney:2018nny},
\begin{equation}
    \epsilon_{H} \simeq \epsilon_{V} = \frac{4}{3 \left(\rm{e}^{\sqrt{\frac{2}{3}} \varphi} -1 \right)^2} \;\;\; \rightarrow \;\;\; r = 16 \epsilon_{V} = \frac{12}{ N^2} \geq 8c^2,
\end{equation}
for a nearly vanishing cosmological constant with $M \simeq \sqrt{3} W_{0} $ or equivalently with $\alpha \simeq 1$. This will reduce the upper limit on $c$ to around $c \lesssim 0.1$ for dSC on the cosmic microwave background scale (see, e.g., \cite{Kinney:2018nny}) and set the upper limit as $c \rightarrow 0$ at the asymptotic limit $\varphi \rightarrow \infty$; thus, the single-field $R+R^2$ Starobinsky model appears to belong to the swampland.

However, the nilpotency of $X$ is imposed into the theory as a Lagrange multiplier. The corresponding equations of motion of constrained superfields are not independent but connected to the locally symmetric action. Due to this, the nature of nilpotent superfields is quite different. Accordingly, the equations of motion of the scalar fields become equivalent with or without removing the nilpotent superfields from the low-energy description of the theory via a unitary gauge imposed on goldstino fermions as follows:
\vspace{-3mm}
\paragraph{Unitary Gauge:}
Equivalently, fulfilling the requirement $X=0$, a gauge fixing can be imposed for the nilpotent superfield with
\begin{equation}
    \chi = 0 \;\;\; \rightarrow \;\;\; X(\chi,F)= \frac{\chi^2}{2F} =0.
\end{equation}
It is a well-known fact that any gauge symmetry requires a gauge fixing condition, and the local SUSY of the SUGRA action becomes broken. That is a gauge fixing condition of the locally supersymmetric chiral matter fermions and not for the gravitino. The gravitino wave operator has nonzero modes and no propagating ghost degrees of freedom in the Euclidean signature \cite{Ferrara:2015tyn,Bergshoeff:2015tra} and it corresponds to a \textit{unitary gauge} \cite{Kallosh:2015sea,Kallosh:2015tea}. That is established and demonstrated at different levels in the context of SUGRA; 

At the level of the equations of motion by combining them algebraically and arranging the constraint via a Lagrange multiplier \cite{Bergshoeff:2016psz}. Following that, at the construction level of the algebra and multiplication laws of superconformal action where Lagrange multiplier appears only in the heaviest component and is satisfied by the equations of motion for the constrained superfield \cite{Ferrara:2016een}. Accordingly, the action before or after the constraint results in the same physical field equations provided with
\begin{equation}\label{SeS}
    S_{\text{eff}}(\Psi) = S(\Psi,X).
\end{equation} 
Hence, one can consistently use constrained superfields without causing any loss or change in the physical equations of motion. With this in mind, another proof is presented in \cite{Bandos:2016xyu} at the level of invariance of action (without the use of equations of motion) by considering the Noether identities and corresponding local symmetry parameters. In that proof, the action (schematically) is given as 
\begin{equation}
    S(\Psi,X) = S_{1}(\Psi) + S_{2}(\Psi,X).
\end{equation}
Here, $S_{1}(\Psi)$ is an action with locally symmetric fields, and $S_{2}(\Psi,X)$ contains a SUSY-breaking sector with a generic coupling of physical fields to constrained superfields. Then applying the local symmetry (and non-linear) transformations reveals an equivalence for the equations of motion as
\begin{equation}\label{S1eS2}
    \frac{\delta S_{1}(\Psi)}{\delta \Psi} =- \frac{\delta S_{2}(\Psi,X)}{\delta \Psi},
\end{equation}
before or after removing the constrained superfield, where the equations of motion for the constrained superfield are identically satisfied with $\delta S_{2} / \delta X=0$. 

Thus, as the first approach suggests, since the left side of equations of motion (\ref{S1eS2}) or (\ref{SeS}) for scalar fields remain unchanged, the use and extraction of nilpotent superfields are justified to generate an effective action. This is also expected since EFT should only be described using IR degrees of freedom, which we obtain after integrating out the massive UV degrees of freedom. Consequently, this constraint can be successfully applied manually or equivalently with a Lagrange multiplier. On the string theory side, this indicates that the resulting low-energy effective potential obtains a positive contribution through the uplifting $\overline{D3}$ branes interacting with other moduli and bulk geometry. Then, the interaction is turned off by imposing $X=0$ to denote the EFT in a unitary gauge.

On the other hand, the right side of equivalence (\ref{S1eS2}) or (\ref{SeS}) presents us an equally interesting scenario. Since we have equivalent equations of motion for scalar fields, one can extend the validity of the interaction of uplifting $\overline{D3}$ branes to cosmological regimes upon imposing the unitary gauge after taking the derivatives. Another way of saying it: nilpotent superfields will open up the possibility of describing a low-energy EFT without invalidating it. Consequently, nilpotent superfields can be kept in theory before applying the unitary gauge, thus contributing to the potential gradient flow ($\propto \nabla V$). Here, it is essential to state that there are no propagating scalar degrees of freedom of nilpotent superfields. Instead, the contribution comes from the fermionic bilinear, which replaced the scalar part and coupled with the scalar terms in the potential.\footnote{Although this is not the approach of this work, one should note that dual descriptions with scalar degrees of freedom of models with constrained superfields can be provided \cite{Farakos:2021oyo}. Alternatively, the validity of constrained superfields can be extended by integrating out heavy modes without neglecting the derivative terms starting from a linearly realized SUSY, thus obtaining nilpotency constraints on superfields in scalar terms \cite{Aoki:2021nna}.} That is the underlying motivation considered in this approach about the viability of embedding the nilpotent superfields into dSC. Therefore, let's take a step back and examine the potential in more detail.

 At this point, one can denote the SUGRA potential for (super)universal attractor setup (assuming $\alpha \simeq 1$) on flat directions as
\begin{equation}\label{SUASUGRAPOT}
     V\left(\Phi, \bar{\Phi},X,\bar{X}\right)\big|_{\Phi=\bar{\Phi}} =  \left[V_0 + V_1 X + \bar{V}_1 \bar{X} + V_2 X \bar{X} + \dots \right] \mathrm{e}^{K\left(\Phi, \bar{\Phi},X,\bar{X}\right) \big|_{\Phi=\bar{\Phi}}}.
\end{equation}
Here, the SUGRA potential contains parts of the scalar potential with canonically normalized real field $\varphi$, and higher order terms in nilpotent superfields were omitted due to the nilpotency constraint $X^2=0$. The zeroth order term gives rise to the $R+R^2$ Starobinsky potential as
\begin{equation}\label{V0}
    V_0 \equiv V_{R+R^2}(\varphi) =  M^2 \left(1 - \rm{e}^{- \sqrt{\frac{2}{3}} \varphi} \right)^2,
\end{equation}
and additional mixed components coupled to nilpotent superfields are given with
\begin{equation}
    V_1(\varphi) = \frac{M^2}{\sqrt{3}}\left[4-\left(1+i\right)\left(\mathrm{e}^{\sqrt{\frac{2}{3}}\varphi} + 3 \mathrm{e}^{-\sqrt{\frac{2}{3}}\varphi}\right) \right], \;\;\; V_2(\varphi) = \frac{M^2}{3} \left(17 \mathrm{e}^{\sqrt{\frac{8}{3}}\varphi} - 14 \mathrm{e}^{\sqrt{\frac{2}{3}}\varphi} + 19 \right)
\end{equation}
upon applying the conformal transformation (\ref{UAmaincomp}) with the plus sign. Note here that the F-terms are denoted by
\begin{equation}
    \mathcal{D}_{\Psi} W \big|_{X=\bar{X}=0} = 0
    , \;\;\; \mathcal{D}_{X} W \big|_{X=\bar{X}=0} =M - M(1+i) \mathrm{e}^{ \sqrt{\frac{2}{3}}\varphi}
\end{equation}
or
\begin{gather}
    \mathcal{D}_{\Psi} W  = \frac{X \bar{X} M }{3} \mathrm{e}^{\sqrt{\frac{8}{3}}\varphi} - \frac{X M}{\sqrt{3}} (1+i) \mathrm{e}^{ \sqrt{\frac{2}{3}}\varphi} + \mathcal{O}(X^2), \\
     \mathcal{D}_{X} W = X \bar{X} M \mathrm{e}^{\sqrt{\frac{8}{3}}\varphi} + \frac{\bar{X} M}{\sqrt{3}} \mathrm{e}^{\sqrt{\frac{8}{3}}\varphi} - X \sqrt{3} M (1+i) \mathrm{e}^{\sqrt{\frac{2}{3}}\varphi} +
    M - M(1+i) \mathrm{e}^{ \sqrt{\frac{2}{3}}\varphi}
\end{gather}
according to the possible scenarios of the left and right side of the equivalences, (\ref{S1eS2}) or (\ref{SeS}). 

For the initial scenario, it is explicitly seen that SUSY is spontaneously and purely broken along the $X$ direction leading to a stable dS vacuum ($\alpha \gtrsim 1$) at the minimum of the potential with $\mathcal{D}_{X} W = - i M$. That term precisely leads to the uplifting contribution for the scalar potential and is obtained by means of linear nilpotent coupling terms in the Kähler potential. Following that, one obtains the scalar potential (\ref{V0}) and model falls into the swampland.

As for the latter scenario, nilpotent superfields are allowed to be held until their derivatives are taken. Following that, one can state the dSC in $\mathcal{N}=1$, $D=4$ SUGRA framework for a positive potential as
\begin{equation}\label{dSConjecture}
   \epsilon_{V}= \frac{g^{I \bar{J}} \partial_{I} V  \partial_{\bar{J}} V}{V^{2}}\bigg|_{X=\bar{X}=0} \geq \frac{1}{2} c^{2}
\end{equation}
where $g^{I \bar{J}}$ is the inverse Kähler metric $K^{I \bar{J}}$. Indices $I,J$ sums over the canonical superfield $\Psi$ and the nilpotent $X$ for the (super)universal attractor mechanism constructed with nilpotent Kähler corrections. Then, for the dSC, the non-vanishing terms of the inverse Kähler metric become $K^{\Psi \bar{\Psi}} = 1$ and $K^{X \bar{X}} = \mathrm{e}^{-\frac{4}{\sqrt{6}} \varphi }$ provided with $\Omega(\phi) = \rm{e}^{ \sqrt{\frac{2}{3}} \varphi}$.
\begin{figure}[htb]
\centering
\includegraphics[width=3.8in]{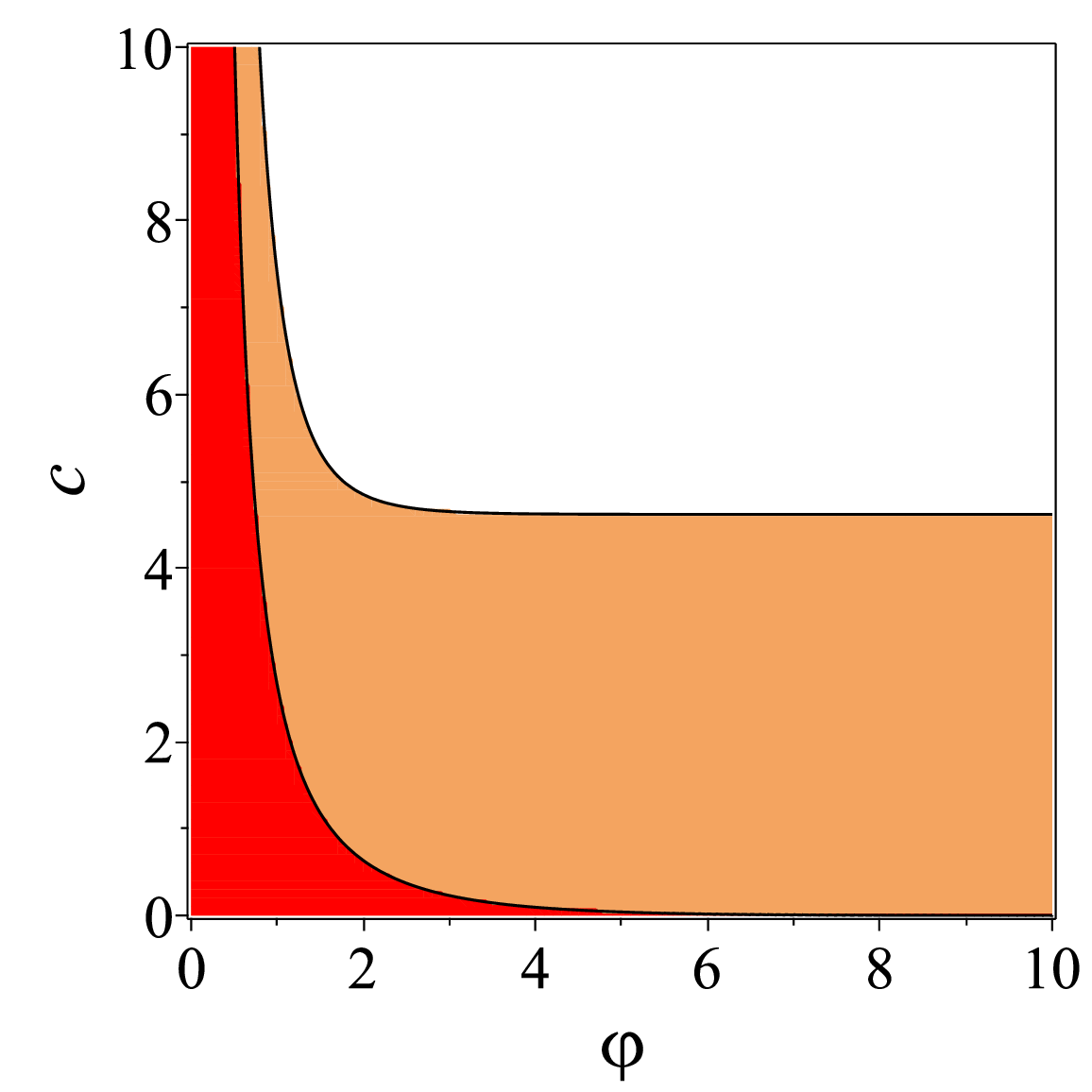}
\caption{\label{F3} Parameter space of $c$ and $\varphi$ for the dSC. The orange(red) area shows the string landscape with(without, i.e. $K^{X \bar{X}} =0$) nilpotent superfield contributions, and the remainder is swampland.
}
\end{figure}
That lead to the dSC inequality (\ref{dSConjecture}) on flat directions as
\begin{equation}\label{dSCcanonical}
     \frac{8 \left(8 - 50 \,{\mathrm e}^{-  \sqrt{6}\varphi}+19 \,{\mathrm e}^{- \sqrt{\frac{32}{3}}\varphi}+60 \,{\mathrm e}^{- \sqrt{\frac{8}{3}}\varphi}-32 \,{\mathrm e}^{- \sqrt{\frac{2}{3}}\varphi}\right)}{3 \left(\mathrm {e}^{- \sqrt{\frac{2}{3}}\varphi}-1\right)^{4}} \gtrsim c^2,
\end{equation}
for a nearly vanishing cosmological constant. As in the stabilization of masses, the corresponding inequality depends on the evolution of the canonically normalized real field $\varphi$, therefore the bound on the parameter $c$ also evolves. That is shown in figure \ref{F3}. That parameter can be bounded as small as $\epsilon_{V}=32/3 \gtrsim c^2 /2$ in the asymptotic limit $\varphi \rightarrow \infty$. Note that, one can also denote the dSC with non-canonical fields as
\begin{equation}
     \frac{8 \left(8 \Omega^{4}-32 \Omega^{3}+60 \Omega^{2}-50 \Omega +19\right)}{3 \left(\Omega -1\right)^{4}} \gtrsim c^2
\end{equation}
and the same bound $\epsilon_{V}=32/3 \gtrsim c^2 /2$ appears when $\Omega \rightarrow \infty$. Hence, the (super)universal attractors are in the string landscape for the dSC provided with $4.61 \gtrsim c \sim \mathcal{O}(1)$. In this case, however, it is necessary to note that the viability of inflation depends on the imposition $\dot{X}=0$, where the Hubble slow roll parameter becomes
\begin{equation}\label{epsilonHstatic}
    \epsilon_{H}\big|_{\dot{X}=0} = \frac{\dot{\varphi}^2}{2 H^2} \simeq \frac{1}{2} \left( \frac{\partial_{\varphi} V}{V}\right)^2< 1,
\end{equation}
thus indicating that nilpotent superfields do not evolve over cosmic time. 

Regardless, the observed difference between $\epsilon_H$ and $\epsilon_V$, as well as the viability of inflation is still in question. In the remainder of this section, we will analyze these scenarios depending on the nature of nilpotent superfields, as well as discuss and offer a way out of that dilemma.
\vspace{-3mm}
\paragraph{Viability of inflation:}
For the first scenario (left-hand side of \ref{S1eS2}), the remaining scalar fields of the theory ($\varphi, \beta$) are stable, and the equation of motion of the inflaton field $\varphi$ is satisfied with $\delta S(\varphi) / \delta \varphi = 0$. In particular, one obtains an attractor solution for the observed parameters of inflation, i.e., $n_s$ and $r$, and inflation is viable with $\epsilon_H \simeq \epsilon_V < 1$. There are no observed differences for $\epsilon_V$ and $\epsilon_H$ as expected for a single-field inflation scenario, indicating that this scenario is in the swampland.

For the second scenario (right-hand side of \ref{S1eS2}), after imposing the equations of motion of the nilpotent superfield with $X^2 =0$, one still has $X$ terms remaining in the potential (\ref{SUASUGRAPOT}), coupled with the inflaton field $\varphi$.\footnote{Even after imposing the equations of motion for $X$ and $\bar{X}$, the potential still depends on these fields. Therefore, the action considered is not fully on-shell. Instead, it is an effective action that takes into account the classical solutions for $X$ and $\bar{X}$, which satisfy their respective equations of motion $X^2=0$, $\bar{X}^2=0$. However, one still needs to take into account the quantum fluctuations of these superfields around their classical solutions. The reason is that even when the vev of a field is zero, its quantum fluctuations can still contribute to physical observables. In QFT, physical observables are calculated by integrating over all possible configurations of the fields, including their quantum fluctuations. They can contribute to the effective potential through loop diagrams, which involve virtual particles propagating off-shell.} Thus, one obtains $\epsilon_V > 1$, including the gradient $\partial_X V$, and the scenario is in the landscape. But it does not necessarily mean that inflation is viable for that scenario. In fact, the nilpotent superfields play an important role in this calculation, as they can affect the vacuum structure and the stability of the theory. Accordingly, one should follow a careful analysis for different definitions of nilpotent superfields. Therefore, let us analyze the three main cases where the nilpotent superfield $X$ is inflaton, time-independent, or time-dependent, respectively.
\begin{itemize}
    \item[\textbf{i.}] Assuming that one can treat the remaining nilpotent superfield as a separate inflaton field under the slow-roll approximations, results in $\epsilon_V \simeq \epsilon_H > 1$, rendering inflation non-viable. However, an inflaton field must have a non-zero vev during inflation to provide the necessary energy density for the inflationary expansion. As the nilpotent superfield has a zero vev, it cannot be considered as a separate inflaton field in a slow-roll inflation scenario.
\end{itemize}

If a nilpotent superfield is coupled to inflaton field, it is possible for the combination of the two fields to act as an effective inflaton field. In this case the inflaton field would be responsible for driving inflation and would have a non-zero potential energy, while the nilpotent superfield would provide a correction to the inflation potential and/or introduce fermionic degrees of freedom.

\begin{itemize}
    \item[\textbf{ii.}] Treating the remaining nilpotent superfield as a fermionic bilinear and assuming it is time-independent ($\dot{X}=0$), the value of the potential will depend on the inflaton field (and any other fields in the theory). In other words, while the fermionic bilinear does not have a kinetic term and therefore does not have a propagating degree of freedom, it can still appear in the potential term of the Lagrangian and contribute to the equations of motion through the potential. Therefore, when considering the equations of motion for inflation, they should be calculated in the direction of the inflaton field only. The equations of motion for a time-independent nilpotent superfield (coupled to an inflaton) will be trivially satisfied due to time independence, i.e.
\begin{equation}
\frac{\delta S(\varphi, X)}{\delta X}=\ddot{X}+3 H \dot{X}+\partial_X V(\varphi, X)=0 \quad \rightarrow \quad \partial_X V(\varphi, X)=0,
\end{equation}
and become identically zero. This means that the potential is already at an extremum with respect to the nilpotent superfield, at that particular value. This could be a minimum, maximum or a saddle point depending on the curvature of the potential in that direction (in our case, it is the minimum with $\Lambda_{dS}$ for $\alpha \gtrsim 1$).
Thus, in the case where $X$ is time-independent, there are no observable differences and one gets $\epsilon_V \simeq \epsilon_H <1$. This means that for the equations of motion in the inflaton direction, this case coincides with the l.h.s. of (\ref{S1eS2}) where $X$ is set to zero.
\end{itemize}

\begin{itemize}
    \item[\textbf{iii.}] For a time-dependent nilpotent superfield, the potential gradient in the $X$ direction can be nonzero while $\dot{X}=0$ at a stationary point of $X$, because it depends on the functional form of the potential $V(\varphi,X,\bar{X})$. At that stationary point, the equation of motion for $X$ again reduces to $\partial V/\partial X = 0$, which also determines the stationary points of the potential energy surface with respect to $X$. However, this does not imply that the potential gradient in the $X$ direction is zero away from those stationary points. The potential gradient in the $X$ direction can be nonzero even if $\partial V/\partial X=0$ at a particular point in the parameter space. This is because the potential gradient includes contributions from all the fields that appear in the potential, not just $X$. Therefore, the specific form of the potential gradient and the dynamics of the system depend on the details of the model and the couplings between the fields. More explicitly, it is suitable to consider the potential (\ref{SUASUGRAPOT}) after imposing $X^2=0$ and applying the conformal transformation (\ref{UAmaincomp}), given by:
\begin{equation}
V(\varphi,X,\bar{X}) = V_{0}(\varphi) + V_{1}(\varphi) X + \bar{V_{1}(\varphi)} \bar{X} + V_{2}(\varphi) X \bar{X}.
\end{equation}
To find stationary points for $X$ and $\bar{X}$, one can first calculate the partial derivative of the potential with respect to $X$:
\begin{equation}
\frac{\partial V}{\partial X} = V_{1}(\varphi) + V_{2}(\varphi)\bar{X}.
\end{equation}
Next, $\partial V/\partial X$ can be set to zero to find the stationary points of the potential energy surface with respect to $X$:
\begin{equation}
\frac{\partial V}{\partial X} = 0 \quad \Rightarrow \quad V_{1}(\varphi) + V_{2}(\varphi)\bar{X} = 0.
\end{equation}
Solving for $\bar{X}$ gives:
\begin{equation}
\bar{X} = -\frac{V_{1}(\varphi)}{V_{2}(\varphi)}.
\end{equation}
At this stationary point, the potential gradient in the $X$ direction is zero since $\partial V/\partial X = V_{1}(\varphi) + V_{2}(\varphi)\bar{X} = 0$.
However, if one considers points away from this stationary point, the potential gradient in the $X$ direction can be nonzero. (Note that both stationary points for $X$ and $\bar{X}$ will change as the inflaton field $\varphi$ evolves during inflation.) For example, if $\bar{X} = 0$ is taken after taking derivatives, one obtains:
\begin{equation}
\frac{\partial V}{\partial X} = V_{1}(\varphi),
\end{equation}
which is nonzero in general, unless $V_{1}(\varphi)=0$. Therefore, the potential gradient in the $X$ direction can be nonzero even if $\dot{X} = 0$ at that stationary point, and it depends on the couplings between the inflaton field and nilpotent superfields. This indicates that the equations of motion for the remaining time-independent nilpotent superfields coincide with those for the remaining time-dependent nilpotent superfields at the stationary points where $\dot{X}=0$. At that point, the equations of motion for both cases are simultaneously satisfied and one gets $\epsilon_V \simeq \epsilon_H < 1$. Elsewhere, away from those stationary points, one can get non-zero contributions to the potential gradient $\partial V/\partial X$ for time-dependent nilpotent superfields while obtaining $\epsilon_V > 1$. Consequently, for the equations of motion in the inflaton direction, this case coincides with the r.h.s. of (\ref{S1eS2}), where $X$ is allowed to be kept in the action until the derivatives are taken.
\end{itemize}
\vspace{-3mm}
\paragraph{Observed difference between $\epsilon_H$ and $\epsilon_V$:} For the cases analyzed above, one obtains equivalence for the equations of motion in the $X$ direction only at stationary points. This can also be seen in the context of string theory by considering the example of an $\overline{D3}$ brane placed at the tip of a warped Calabi-Yau throat in a static configuration, which does not evolve with cosmic time. In that sense, one can impose $\dot{X}=0$ as a static solution in the energy conservation equation (\ref{EC}):
\begin{equation}
\left[\ddot{X}+3 H \dot{X}+\partial_X V(\varphi, X)\right] \dot{X}=0 \quad \rightarrow \quad\left[0+0+\partial_X V(\varphi, X)\right] 0=0,
\end{equation}
as shown in the appendix for a general canonical multi-field case. Hence, both $\partial_X V = 0$ and $\partial_X V \neq 0$ are allowed when the energy conservation equation is identically zero, where one obtains (\ref{epsilonHstatic}) for the Hubble slow-roll parameter with $\epsilon_H < 1$. Consequently, the energy conservation equation becomes identically zero for the $X$ direction while the number of equations for scalar fields remains the same. This is also implied similarly by (\ref{S1eS2}) for every value of the inflaton field $\varphi$:
\begin{equation}\label{S1S2eomEQU}
\frac{\delta S_{1}(\varphi)}{\delta \varphi} =- \frac{\delta S_{2}(\varphi,X)}{\delta \varphi}.
\end{equation}

However, in order to obtain a simultaneous observable difference between $\epsilon_H$ and $\epsilon_V$, it is necessary to consider the r.h.s. of this equivalence for a viable inflation period (since $X$ has a zero vev in any case considered above and cannot be treated as a separate inflaton field for slow-roll inflation), while the r.h.s. of the equivalence represents the potential gradient with a time-evolving nilpotent superfield $(\dot{X} \neq 0)$, thus describing the string landscape. For this reason, this approach relies on the equivalence of the equations of motion of the inflaton field $\varphi$, where one side presents a viable inflaton scenario and the other side presents a theory in the landscape, as follows:
\begin{equation}
\epsilon_{H}\big|_{\text{l.h.s.}} \ll 1 \;\;\; \text{and} \;\;\; \epsilon_{V}\big|_{\text{r.h.s.}} \geq  \mathcal{O}(1).
\end{equation}
Most importantly, unlike in the traditional sense for a multi-field case as denoted in \cite{Achucarro:2018vey}, the observable difference does not appear separately only on the r.h.s. or only on the l.h.s. of the equivalence (\ref{S1S2eomEQU}).

On the l.h.s. of the equivalence, the potential is solely determined by the inflaton field $\varphi$, as the nilpotent superfields have zero vev and no propagating degrees of freedom. These auxiliary fields are introduced to stabilize the vacuum energy and enable a viable inflationary scenario with an uplifted potential, yet their presence does not give rise to independent degrees of freedom. Thus, the gradient with respect to $\varphi$ is the only relevant quantity that governs the dynamics of the system and the stability conditions for inflation, as demonstrated in section \ref{Sec3.3}.

On the r.h.s., the potential depends on both the inflaton field and the nilpotent superfields $X$ and $\bar{X}$. When these fields have time-dependent dynamical degrees of freedom, their gradients may not vanish away from the stationary points when $\dot{X}=0$. Consequently, these terms introduce novel contributions to the equations of motion for the $X$ direction, necessitating the analysis of the system's stability as a whole. Since the inflaton field and the nilpotent superfields are coupled, their stability conditions are interconnected, and the stability of one field may influence that of the others. In particular, the time-dependence of the nilpotent superfields can generate instabilities on the r.h.s. of the equivalence (\ref{S1S2eomEQU}), requiring a thorough examination. Failing to do so could impact the evolution of the system and potentially invalidate the inflationary scenario. Therefore, it is crucial to consider both the time-dependence of the nilpotent superfields and their gradients when investigating the stability of the system.


\section{Conclusions}\label{Sec4}
\hspace{4mm}
In this study, we build an EFT for inflation with spontaneously broken SUSY without generating any supersymmetric AdS vacua. To do this, we combined the tools offered for the nilpotent SUGRA, which effectively captures the physics of $\overline{D3}$ branes, and the universal attractor mechanism for inflation. For all calculations, we used the superpotential known to describe a dS phase in SUGRA:
\begin{equation}
    W= W_0 + M X,
\end{equation}
where $X$ corresponds to the nilpotent superfield, $W_0$ is the gravitino mass, and $M$ parameterizes the $\overline{D3}$ brane contribution. Given the sequestered inflation conditions, we denoted that the only non-vanishing F-term becomes
\begin{equation}  \mathcal{D}_{X} W = M + g(\Phi,\bar{\Phi}) W_{0},
\end{equation}
for the most general form of the Kähler potential in the context of nilpotent SUGRA. Hence following that, we showed that the linear coupling $g(\Phi,\bar{\Phi})$ plays a crucial role in inflation, spontaneous SUSY breaking, and obtaining a dS vacua at the minimum of the potential where $\Phi$ represents a chiral superfield. After that, to avoid any fine-tuning and provide a more natural basis, we expressed the final Kähler potential as
\begin{equation}
    K\left(\Phi, \bar{\Phi}, X,\bar{X} \right)=K_{\text{main}}\left(\Phi, \bar{\Phi}\right) + \frac{1}{6 \alpha} g(\Phi) \bar{g}(\bar{\Phi}) X \bar{X} + g(\Phi) X + \bar{g}(\bar{\Phi}) \bar{X},
\end{equation}
which captures the same pattern with universal attractors for inflation with a single parameter 
\begin{equation}
    g(\Phi) = -\sqrt{3 \alpha} \left( \Omega(\Phi) + i \Omega(\Phi)   \right) 
\end{equation}
where $\Omega(\Phi)$ is the main component of that mechanism. At the same time, that is equivalent to the conformal transformation function that switches the Jordan frame to the Einstein frame in flat directions. The imaginary portion $i \Omega(\Phi)$ fulfills exactly the required uplifting contribution for the potential and one obtains dS vacua for
\begin{equation}
    \sqrt{\alpha} = \left|\frac{M}{\sqrt{3} W_0} \right|  > 1.
\end{equation}
Here, the main part of the Kähler potential, $K_{\text{main}}\left(\Phi, \bar{\Phi} \right)$, is only responsible for producing the expected kinetic term and consequently the conformal transformation function in flat directions. It does not affect the scalar potential with non-canonical fields for the sequestered inflation mechanism. In that regard, we showed that various Kähler potentials could produce the same kinetic term in flat directions without disordering the scalar potential up to a Kähler transformation. 

The resulting SUGRA potential led to the scalar formulation of $R+R^2$ Starobinsky potential, and a residual term for the cosmological constant as 
\begin{equation}
    \Lambda_{dS} = M^2\left(1-\frac{1}{\alpha}\right) = M^2 - 3 W_{0}^2 >0,
\end{equation}
at the minimum of the scalar potential. Furthermore, the $\overline{D3}$ brane contribution $M$ was determined by the observed value of the power spectrum of curvature amplitude, hence predicting an equally small gravitino mass $W_0 \simeq \pm 6.36 \times 10^{-6}$ for an almost vanishing cosmological constant and for $N\simeq 56.2$ e-folds. Also, as one of the most notable features of this approach, we showed that the potential reduces to a dS or Minkowski state when the $\overline{D3}$ brane contribution is excluded by taking the $M\rightarrow0$ limit (for $\alpha \geq 1/2$); otherwise, SUSY remains spontaneously broken at all times. Thus, for this approach, one achieves an uplifting to dS vacua without producing or the necessity of any supersymmetric AdS vacua.

Following that, we denoted that one can promote the parameter $\alpha$ as an overall multiplier to the total Kähler potential as $K^{\prime} = \alpha K$ without changing the structure of the nilpotent part of the Kähler potential for the linear coupling $ g(\Phi) = -\sqrt{3 }\alpha \left( \Omega(\Phi) + i \Omega(\Phi)   \right) $. That allowed us to capture a unified region of the inflationary attractors with \cite{Galante:2014ifa}:
\begin{equation}
    \alpha = 1 + \frac{1}{6 \xi},
\end{equation}
where $\xi>0$ at all times to avoid negative kinetic terms in the Jordan frame. For that region, one obtains a dS vacuum with an almost vanishing cosmological constant in the large $\xi$ limit and a Minkowski vacuum in the limit, $\xi \rightarrow \infty$. Accordingly, obtaining AdS vacua stays out of the picture. 

As we continue with the (super)universal attractors, we demonstrated that all real canonically normalized fields are in the correct hierarchical order for a successful period of single field slow-roll inflation, and the moduli are stable. Finally, we embedded the nilpotent superfields into the dSC in an $\mathcal{N}=1$, $D=4$ SUGRA background. That conjecture usually rules out single field slow-roll inflation, yet the presence of nilpotent superfields allowed us to expand the consistent EFT space for that conjecture. We showed that the bound on the parameter $c\sim \mathcal{O}(1)$ of the dSC dynamically evolves during the inflation period in case nilpotent superfields are allowed to be included to the potential gradient. The smallest bound appears on the asymptotic limit $\Omega \rightarrow \infty$ (or $\varphi \rightarrow \infty$) as $4.61 \gtrsim c $; hence the (super)universal attractors provide an elegant cosmological picture for the swampland while effectively realizing dS vacua from string theory.

\section*{Acknowledgment}
The author acknowledges Georges Obied and Alek Bedroya for the stimulating discussions and Mehmet Ozkan for valuable suggestions and comments. The author also would like to thank Renata Kallosh for valuable clarifications and discussions towards completing this work. Finally, the author would like to thank Cumrun Vafa and the hospitality of Harvard University, where this work was done. O.G. also acknowledges the support by the Outstanding Young Scientist Award of the Turkish Academy of Sciences (TUBA-GEBIP) and TUBITAK 2214-A grant 1059B142100284.

\appendix
\section{On the Equivalence and Transition in Multi-field Theories}\label{Appendix}
\hspace{4mm}
In the first part of this appendix, we denote a general canonical case with multiple inflaton fields and show how the approximate equivalence between the Hubble slow-roll parameter and potential slow-roll parameter arises. By doing so, we make it clear that this approximate equivalence generally holds for this class of slow-roll inflation models, otherwise would not hold. In the second part, we highlight a transition between the main generating sectors, Kähler potential, and superpotential. Then, we show a transformation based on the nilpotency condition and simultaneous symmetries, which are preserved under a suitable field redefinition. Accordingly, after this procedure, we show that the nilpotent part of the most general Kähler potential can be described without explicit dependence/couplings on moduli fields, i.e., $\Phi$.

\subsection{(In)Equivalence of the Slow-roll Criteria and dSC in Multi-field Theories}\label{A.1}
\hspace{4mm}
A general definition of inflation can be given in terms of the Hubble parameter $H\equiv \dot{a}/a$ and the scale factor $a(t)$, with
\begin{equation}
    \frac{d}{dt}\left(\frac{1}{a H}\right)=\frac{1}{a}\left(-\frac{\dot{H}}{H^2}-1\right)=-\frac{\ddot{a}}{a} < 0,
\end{equation}
as a decreasing Hubble radius or equivalently as an accelerated scale factor parameter ($\ddot{a}>0$) is needed for an inflationary epoch in the early universe. Then accordingly, it is also convenient to define inflation with the condition
\begin{equation}\label{HEP}
    \epsilon_{H} \equiv -\frac{\dot{H}}{H^2} <1.
\end{equation}
That is a general condition for inflation and independent of the nature of scalar fields (if any) exist in the theory. Keeping this in mind, it is possible to build various inflationary models with multiple fields \cite{GrootNibbelink:2001qt,Bassett:2005xm,Malik:2008im}, and one of them is multi-field slow-roll inflation, where every field of the theory is \textit{approximated under the same conditions} according to the slow-roll approximations.\footnote{Note that fixing an inflation trajectory in which only a single scalar field lives may effectively result in a single-field inflation scenario. In this perspective, a detailed and intensive effort has been made to capture the single-field effective inflation models from the multi-field theories whose various aspects are explored in \cite{Chen:2009zp,Achucarro:2010da,Achucarro:2011yc,Shiu:2011qw,Avgoustidis:2011em,Achucarro:2012sm,Cespedes:2012hu,Avgoustidis:2012yc,Chen:2012ge,Pi:2012gf,Gao:2012uq,Achucarro:2012yr,Collins:2012nq,Burgess:2012dz,Gwyn:2012mw,Noumi:2012vr,Dimastrogiovanni:2012st,Bartolo:2013exa,Garcia-Saenz:2018vqf,Durakovic:2019kqq,Pinol:2020kvw,Romano:2022hli}.}

Let us, for simplicity, consider the Lagrangian (\ref{EFL}) with canonical kinetic terms and subsequently redefine $\Phi \rightarrow \phi^{a}$ to include multiple fields with $a=\left(1,2,\ldots\right)$ and $\phi=\left(\phi^1,\phi^2, \ldots\right)$ where every index labels a different possible field for the (classical) action:
\begin{equation}
    S=\int d^4 x \mathcal{L}_{\text{E}}= \int d^4 x \sqrt{-g}\left[\frac{R}{2}-\frac{1}{2} g^{\mu \nu} \partial_\mu \phi^a \partial_\nu \phi_a-V(\phi)\right].
\end{equation}
Here we take $M_{pl}\equiv(8 \pi G)^{-1/2} =1$ and use dummy index notation to show the sum over all fields for the kinetic term. For this action, the variation with respect to space-time metric tensor $\delta S / \delta g_{\mu\nu}=0$, defined in terms of FLRW metric 
\begin{equation}
    d s^2=-d t^2 +a^2(t) d \mathbf{x}^2 ,
\end{equation}
leads to Friedmann equations (background dynamics),
\begin{equation}\label{FE1}
3 H^2 =\rho=\frac{\dot{\phi}^a \dot{\phi}_a}{2}+V\left(\phi \right),
\end{equation}
\begin{equation}\label{FE2}
    \dot{H} =-\frac{\rho + p}{2}=-\frac{\dot{\phi}^a \dot{\phi}_a}{2} 
\end{equation}
where the total energy density ($\rho=\dot{\phi}^a \dot{\phi}_a /2 +V$) and pressure ($p=\dot{\phi}^a \dot{\phi}_a /2 -V$) are determined with the corresponding stress-energy tensor considering Einstein equations $G_{\mu\nu} = T_{\mu\nu}$. Once again considering the Einstein equations, one of the Bianchi identities lead to the energy conservation equation as
\begin{equation}\label{EC}
    \dot{\rho} + 3 H (\rho + p) =  0 \quad \rightarrow \quad  \left[ \ddot{\phi^a} + 3 H \dot{\phi^a} + \partial_{\phi^a} V(\phi) \right] \dot{\phi_a} = 0
\end{equation}
where inside the parenthesis corresponds to Klein-Gordon equation and one can equivalently derive it as equations of motion for $\phi^a$ by taking the variation $\delta S / \delta \phi^a = 0$. As moving on, before imposing any slow-roll criteria, one can rearrange the Hubble slow-roll parameter using the Friedmann equation (\ref{FE2}) with (\ref{HEP}) and denote
\begin{equation}\label{HubbleSRP}
    \epsilon_{H} = \frac{\dot{\phi}^a \dot{\phi}_a}{2 H^2},
\end{equation}
where the collective summation is required to be smaller then unity during the inflation epoch for every field of the theory with $\epsilon_{H} = \epsilon_{H_1} + \epsilon_{H_2} + \ldots <1$.
\vspace{-3mm}
\paragraph{Equivalence ($\epsilon_{H}\simeq \epsilon_{V}$):}
Then, in order to obtain a sufficiently long period of inflation with $\epsilon_{H} \ll 1$, one can consider the slow-roll condition and the time-derivative of this condition (in absolute value) as
\begin{equation}\label{SRCriteria}
    \frac{\dot{\phi}^a \dot{\phi}_a}{2} \ll V(\phi), \quad \big|\ddot{\phi^a}\dot{\phi_a}\big| \ll \big|\partial_{\phi^a} V(\phi)\dot{\phi_a}\big|.
\end{equation}
Following those conditions, one obtains the slow-roll approximations using (\ref{FE1}) and (\ref{EC}) as
\begin{equation}\label{SRapprox}
    3H^2 \simeq V(\phi), \quad 3 H \dot{\phi^a}\dot{\phi_a} \simeq - \partial_{\phi^a} V(\phi)\dot{\phi_a}.
\end{equation}
Inserting these two approximate equations into (\ref{HubbleSRP}) by considering every scalar field (inflaton) of the theory fulfills them \textit{separately}, one can denote
\begin{equation}\label{appEqEpsilon}
    \epsilon_{H} = \frac{\dot{\phi}^a \dot{\phi}_a}{2 H^2}\simeq  \frac{1}{2}\left[\left(\frac{\partial_{\phi^1} V}{V} \right)^2 +\left(\frac{\partial_{\phi^2} V}{V} \right)^2 + \ldots \right] = \frac{1}{2} \left(\frac{\nabla V}{V} \right)^2 = \epsilon_{V}
\end{equation}
as a first-order result of the imposed approximations. Accordingly, in case the approximations, (\ref{SRapprox}) are valid for all inflaton fields, then the distinction between those parameters can be ignored and consequently the gradient flow of the potential with respect to all fields ($\propto \nabla V$), as stated for the dSC, becomes approximately equivalent to the Hubble slow-roll parameter with $\epsilon_{H} \simeq \epsilon_V$.  Thus, only in that class of models (including the usual single-field inflation), inflation directly contradicts with the dSC since $\epsilon_V \ll 1$ during the slow-roll inflation, and the dSC requires $\epsilon_V \geq  \mathcal{O}(1)$. However, in case the slow-roll criteria (\ref{SRCriteria}) does not hold for every field, then the approximate equivalence is broken down, i.e., $\epsilon_{H} \not\simeq \epsilon_V$, and thus opens a possibility of evaluating them separately.

\subsection{An Equivalence of Two Main Generating Sectors}\label{A.2}
\hspace{4mm}
Another attractive property of the nilpotent SUGRA approach is that one can establish equivalence with generic superpotential models (or vice-versa) with a simple field redefinition, i.e., a holomorphic change of the Kähler manifold coordinates, corresponding to a nilpotent superfield. Equivalently, this procedure presents a transition between models with curved ($K_{X\bar{X}} \neq 1$) and flat ($K_{X^{\prime}\bar{X^{\prime}}} = 1$) geometries of nilpotent superfields, as also remarked for $\overline{D3}$-induced geometric inflation scenarios \cite{Kallosh:2017wnt}. 

In that respect, the corresponding nilpotent part of the Kähler potential for (super)universal attractors retains the same structure in terms of the linear nilpotent couplings as
\begin{equation}\label{A.12}
    K_{N.C.}=  \frac{1}{6\alpha} g \bar{g} X \bar{X} + g X + \bar{g} \bar{X},
\end{equation}
which is also the case for the unification limit (\ref{UniLim}). Hence in general, the reality condition $K = \bar{K}$ on the Kähler potential (\ref{Knc}) or (\ref{KncAlpha}) implies that there should be a \textit{simultaneous} symmetry on the linear couplings and the nilpotent superfields at the same time as; \begin{equation}
    g(\Phi) \rightarrow \bar{g}(\bar{\Phi}) \;\;\; \text{and }\;\;\; X \rightarrow \bar{X}.
\end{equation}
Then one can establish a simple equivalence with the generic superpotential cases by defining
\begin{equation}
    g(\Phi) X  \rightarrow X^{\prime}
\end{equation}
where $X^{\prime}$ shares the same properties with the nilpotent superfield $X$ since it also satisfies the same constraint as 
\begin{equation}\label{NilpotencyCond}
    (X^{\prime})^2 \equiv g(\Phi)^2 X^2 = 0.
\end{equation}
Consequently, $X^{\prime}$ removes the explicit $\Phi$ dependence of the nilpotent Kähler corrections and diverts it to the superpotential with a generic function of $\Phi$:
\begin{equation}
    K^{\prime}=K_{\text{main}} +\frac{1}{6 \alpha} X^{\prime} \bar{X}^{\prime} + X^{\prime} + \bar{X}^{\prime} \;\;\; \text{and} \;\;\; W = W_{\text{main}} + W_0 + g^{-1} M X^{\prime}.
\end{equation}

More generally, this transition also applies to the most general form of the Kähler potential with nilpotent corrections (\ref{MostGenKahler}) since the nilpotency condition (\ref{NilpotencyCond}) is also satisfied for the linear non-holomorphic generic coupling terms $g(\Phi,\bar{\Phi}) X$. However, an additional condition arises that the mapping of the bilinear term $X \bar{X}$ is required to be proportional to 
\begin{equation}
    f(\Phi, \bar{\Phi})  \propto g(\Phi,\bar{\Phi}) \bar{g}(\Phi,\bar{\Phi})
\end{equation}
with linear couplings. Hence in terms of superfields, the structure of the Kähler potential should remain the same as (\ref{A.12}) regardless of different constant multipliers. Also, note that the $F$-term SUGRA potential (\ref{MostmostGenKah}) contains differential terms as
\begin{equation}
    \frac{\partial_{\Phi}\bar{g}(\Phi,\bar{\Phi}) \partial_{\bar{\Phi}} g(\Phi,\bar{\Phi})}{\partial_{\Phi}\partial_{\bar{\Phi}} K_{\text{main}}(\Phi,\bar{\Phi})},
\end{equation}
that will reappear with non-holomorphic coupling terms, which could drastically change the evolution of the theory. Therefore, it would be very interesting to see further applications of the non-holomorphic terms of nilpotent linear couplings to cosmology, and is left for future research.



		\providecommand{\href}[2]{#2}\begingroup\raggedright\endgroup
		\newpage

	\end{document}